\documentclass[12pt]{article}
\usepackage{amsmath,amssymb,amsthm,amsxtra,overpic,bbm,bm,epsfig,subfigure}
\usepackage{booktabs}
\usepackage{hyperref}
\usepackage{mathrsfs}
\usepackage{graphicx}
\usepackage{multirow}
\usepackage{makecell}
\usepackage{verbatim}
\usepackage{color}
\usepackage{comment}
\usepackage{epstopdf}
\usepackage{float}
\numberwithin{equation}{section}
\usepackage{cite}
\usepackage{hyperref}
\usepackage{url}
\usepackage{slashed,stmaryrd}

\begin{document}
	
	\vspace{0.2cm}
	\begin{center}
		{\Large\bf Non-unitary Leptonic Flavor Mixing and CP Violation \\in Neutrino-antineutrino Oscillations}
	\end{center}
	\vspace{0.2cm}
	
	\begin{center}
		{\bf Yilin Wang}~$^{a,~b}$~\footnote{E-mail: wangyilin@ihep.ac.cn},
		\quad
		{\bf Shun Zhou}~$^{a,~b}$~\footnote{E-mail: zhoush@ihep.ac.cn (corresponding author)}
		\\
		\vspace{0.2cm}
		{\small
			$^a$Institute of High Energy Physics, Chinese Academy of Sciences, Beijing 100049, China\\
			$^b$School of Physical Sciences, University of Chinese Academy of Sciences, Beijing 100049, China}
	\end{center}
	
	\vspace{0.5cm}
	
\begin{abstract}
If massive neutrinos are Majorana particles, then the lepton number should be violated in nature and neutrino-antineutrino oscillations $\nu^{}_\alpha \leftrightarrow \overline{\nu}^{}_\beta$ (for $\alpha, \beta = e, \mu, \tau$) will definitely take place. In the present paper, we study the properties of CP violation in neutrino-antineutrino oscillations with the non-unitary leptonic flavor mixing matrix, which is actually a natural prediction in the canonical seesaw model due to the mixing between light and heavy Majorana neutrinos. The oscillation probabilities $P(\nu^{}_\alpha \to \overline{\nu}^{}_\beta)$ and $P(\overline{\nu}^{}_\alpha \to \nu^{}_\beta)$ are derived, and the CP asymmetries ${\cal A}^{}_{\alpha \beta} \equiv [P(\nu^{}_\alpha \to \overline{\nu}^{}_\beta) - P(\overline{\nu}^{}_\alpha \to \nu^{}_\beta)]/[P(\nu^{}_\alpha \to \overline{\nu}^{}_\beta) + P(\overline{\nu}^{}_\alpha \to \nu^{}_\beta)]$ are also calculated. Taking into account current experimental bounds on the leptonic unitarity violation, we show that the CP asymmetries induced by the non-unitary mixing parameters can significantly deviate from those in the limit of a unitary leptonic flavor mixing.
\end{abstract}

\newpage
\section{Introduction}
\label{sec:intro}

 Neutrino oscillation experiments have provided us with very convincing evidence that neutrinos are actually massive and lepton flavors are significantly mixed~\cite{Zyla:2020zbs, Xing:2019vks}. In order to accommodate tiny neutrino masses, one can naturally extend the Standard Model (SM) by introducing three right-handed neutrino singlets $N^{}_{i{\rm R}}$ (for $i = 1, 2, 3$). After the spontaneous breaking of the SM gauge symmetry, the overall neutrino mass term can be written as~\cite{Xing:2011zza}
 \begin{eqnarray}
 {\cal L}^{}_\nu = - \frac{1}{2} \overline{\left( \begin{matrix} \nu^{}_{\rm L} & N^{\rm C}_{\rm R} \end{matrix} \right)} \left( \begin{matrix} {\bf 0} & M^{}_{\rm D} \cr M^{\rm T}_{\rm D} & M^{}_{\rm R} \end{matrix} \right) \left( \begin{matrix} \nu^{\rm C}_{\rm L} \cr N^{}_{\rm R}\end{matrix} \right) + {\rm h.c.} \; ,
 \label{eq:massterm}
 \end{eqnarray}
where $\nu^{\rm C}_{\rm L} \equiv {\sf C}\overline{\nu^{}_{\rm L}}^{\rm T}$ and $N^{\rm C}_{\rm R} \equiv {\sf C}\overline{N^{}_{\rm R}}^{\rm T}$ with ${\sf C} \equiv {\rm i}\gamma^2 \gamma^0$ stand respectively for the charge conjugates of the left-handed and right-handed neutrino fields, $M^{}_{\rm D}$ for the Dirac neutrino mass matrix, and $M^{}_{\rm R}$ for the Majorana mass matrix of right-handed neutrino singlets. In the flavor basis where the charged-lepton mass matrix $M^{}_l = {\rm Diag}\{m^{}_e, m^{}_\mu, m^{}_\tau\}$ is diagonal with $m^{}_\alpha$ (for $\alpha = e, \mu, \tau$) being the charged-lepton masses, one can diagonalize the $6\times 6$ neutrino mass matrix in Eq.~(\ref{eq:massterm}) by a $6\times 6$ unitary matrix via
\begin{eqnarray}
\left( \begin{matrix} V & R \cr S & U \end{matrix} \right)^\dagger \left( \begin{matrix} {\bf 0} & M^{}_{\rm D} \cr M^{\rm T}_{\rm D} & M^{}_{\rm R} \end{matrix} \right) \left( \begin{matrix} V & R \cr S & U \end{matrix} \right)^* = \left( \begin{matrix} \widehat{M}^{}_\nu & {\bf 0} \cr {\bf 0} & \widehat{M}^{}_{\rm R} \end{matrix} \right) \; ,
\label{eq:unitary}
\end{eqnarray}
where $\widehat{M}^{}_\nu \equiv {\rm Diag}\{m^{}_1, m^{}_2, m^{}_3\}$ and $\widehat{M}^{}_{\rm R} \equiv {\rm Diag}\{M^{}_1, M^{}_2, M^{}_3\}$ with $m^{}_i$ and $M^{}_i$ (for $i = 1, 2, 3$) being the masses of three light and heavy Majorana neutrinos, respectively. Obviously, all the $3\times 3$ matrices $V$, $R$, $S$ and $U$ themselves are not unitary but satisfy the unitarity conditions, such as $V V^\dagger + R R^\dagger = {\bf 1}$ and $V S^\dagger + R U^\dagger = {\bf 0}$. In the mass basis, the charged-current interaction for both light and heavy Majorana neutrinos turns out to be
\begin{eqnarray}
\mathcal{L}_{\rm cc}^{} = \frac{g}{\sqrt{2}} \left[ \overline{\begin{pmatrix} e & \mu & \tau \end{pmatrix}_{\rm L}^{}} V \gamma^\mu_{} \begin{pmatrix}\nu_1^{}\\ \nu_2^{}\\ \nu_3^{}\end{pmatrix}_{\rm L}^{} W_\mu^- + \overline{\begin{pmatrix} e & \mu & \tau \end{pmatrix}_{\rm L}^{}} R \gamma^\mu_{}  \begin{pmatrix} N_1^{}\cr N_2^{} \cr N^{}_3 \end{pmatrix}_{\rm L}^{} W_\mu^-\right] + {\rm h.c.} \,,
\label{cc}
\end{eqnarray}
where the non-unitary matrix $V$ will be involved in the production and detection of light neutrinos, and responsible for the leptonic flavor mixing in the neutrino-neutrino (i.e., $\nu^{}_\alpha \to \nu^{}_\beta$ and $\overline{\nu}^{}_\alpha \to \overline{\nu}^{}_\beta$) and neutrino-antineutrino (i.e., $\nu^{}_\alpha \to \overline{\nu}^{}_\beta$ and $\overline{\nu}^{}_\alpha \to \nu^{}_\beta$) oscillations.

In this canonical type-I seesaw model~\cite{Minkowski, Yanagida, Gellmann, Glashow, Mohapatra}, the effective Majorana mass matrix of three light neutrinos is given by $M^{}_\nu \equiv V \widehat{M}^{}_\nu V^{\rm T} \approx  - M^{}_{\rm D} M^{-1}_{\rm R} M^{\rm T}_{\rm D}$, so the tiny Majorana masses of ordinary neutrinos $\nu^{}_i$ can be attributed to the large masses of heavy Majorana neutrinos $N^{}_i$. If the masses of heavy Majorana neutrinos are around a superhigh-energy scale ${\cal O}(M^{}_{\rm R}) \sim 10^{14}~{\rm GeV}$, then the light Majorana neutrino masses correctly reach the sub-eV level ${\cal O}(M^{}_\nu) \sim 0.1~{\rm eV}$ for ${\cal O}(M^{}_{\rm D}) \sim 10^2~{\rm GeV}$ at the electroweak scale. Consequently, the unitarity violation of the leptonic flavor mixing matrix $V$ is highly suppressed, namely, ${\bf 1} - V V^\dagger = R R^\dagger$ with $R \sim {\cal O}(M^{}_{\rm D} M^{-1}_{\rm R}) \sim 10^{-12}$~\cite{Xing:2005kh}. Generally speaking, the absolute scale of the heavy Majorana neutrino masses cannot be uniquely fixed. As has been pointed out in Ref.~\cite{Pilaftsis:1991ug}, if there exists some symmetry guaranteeing $M^{}_{\rm D} M^{-1}_{\rm R} M^{\rm T}_{\rm D} = {\bf 0}$, then the tiny Majorana neutrino masses of $\nu^{}_i$ are vanishing at the tree level, but they can be radiatively generated. In this case, the light-heavy mixing matrix $R \sim {\cal O}(M^{}_{\rm D} M^{-1}_{\rm R})$ could be as sizable as $10^{-2}$ for ${\cal O}(M^{}_{\rm R}) \sim 10~{\rm TeV}$ or even lower. In such low-scale type-I seesaw models, the heavy Majorana neutrinos are hopefully accessible to the high-energy collider experiments~\cite{Bray:2005wv, Kersten:2007vk, Deppisch:2015qwa, Antusch:2016ejd}, and the resonant leptogenesis mechanism~\cite{Pilaftsis:2003gt} works well to account for the cosmological matter-antimatter asymmetry~\cite{Pilaftsis:2004xx, Pilaftsis:2005rv, Deppisch:2010fr}. On the other hand, the leptonic unitarity violation will receive stringent bounds from electroweak precision data, lepton-flavor-violating decays of charged leptons, and neutrino oscillation experiments~\cite{Antusch:2006vwa, Antusch:2014woa, Escrihuela:2015wra, Fernandez-Martinez:2016lgt}.

In Ref.~\cite{FernandezMartinez:2007ms}, it has been recognized that the non-unitarity of the leptonic flavor mixing matrix $V$ brings in extra sources of CP violation, which can be probed in future long-baseline accelerator neutrino oscillation experiments~\cite{Antusch:2009pm, Blennow:2016jkn, Escrihuela:2016ube, Ellis:2020hus, Forero:2021azc}. In this work, we concentrate on the CP violation induced by the non-unitary flavor mixing matrix $V$ in the neutrino-antineutrino oscillations. The motivation for such an investigation is two-fold. First, since it was suggested by Pontecorvo in 1957~\cite{Pontecorvo:1957cp} that neutrino-antineutrino conversions might occur, there has been great progress in understanding the basic properties of massive neutrinos. Now we know that neutrino-antineutrino oscillations definitely indicate the lepton number violation, and thus take place only if massive neutrinos are Majorana particles~\cite{Majorana:1937vz}. Therefore, it is interesting to examine neutrino-antineutrino oscillations in the type-I seesaw model where neutrinos are indeed Majorana particles and the flavor mixing matrix is intrinsically non-unitary. Second, there exist extensive studies of neutrino-antineutrino oscillations with a unitary flavor mixing matrix~\cite{Bahcall:1978jn, Chang:1980qw, Schechter:1980gk, Li:1981um, Bernabeu:1982vi, Vergados:1985pq, Langacker:1998pv, deGouvea:2002gf, Delepine:2009qg, Xing:2013ty, Xing:2013woa, Kimura:2021juc}. On the one hand, although the oscillation amplitudes are in reality significantly suppressed by the tiny ratios of neutrino masses to neutrino beam energies, the CP asymmetries depend on the Majorana CP-violating phases as well and possess intriguing properties~\cite{Xing:2013ty, Xing:2013woa}. For this reason, we are curious about how different the CP violation with a non-unitary mixing matrix is from that with a unitary one, and how large the deviations can be in light of the latest experimental bounds on unitarity violation. In connection with neutrino-antineutrino oscillations and CP violation, we briefly comment on the oscillations of heavy Majorana neutrinos in the seesaw model~\cite{Antusch:2020pnn} and the resonantly-enhanced CP violation if two heavy Majorana neutrinos become nearly degenerate in mass~\cite{Bray:2007ru}.

The remaining part of this paper is organized as follows. In Sec.~\ref{sec:para}, we give some helpful remarks on the conventional parametrizations of a non-unitary mixing matrix, and clarify the relationship between the Hermitian parametrization and the lower-triangular one. Then, the CP asymmetries for neutrino-antineutrino oscillations with a non-unitary mixing matrix are calculated in Sec.~\ref{sec:CP}, and compared with those in the unitary limit. A brief discussion about neutrino-antineutrino oscillations and CP asymmetries for heavy Majorana neutrinos is also given. We summarize our main results in Sec.~\ref{sec:sum}. Finally, some details about the QR factorization are presented in Appendix~\ref{app:QR}, and the CP asymmetries in neutrino-antineutrino oscillations with a unitary mixing matrix are collected in Appendix~\ref{app:UN}.

\section{Non-unitary Mixing Matrix}
\label{sec:para}

Before calculating the $\nu^{}_\alpha \leftrightarrow \overline{\nu}^{}_\beta$ oscillation probabilities, we first carry out a comparative study of the existing parametrizations for a non-unitary mixing matrix. As is well known~\cite{Antusch:2006vwa, Antusch:2014woa}, the $3\times 3$ non-unitary mixing matrix $V$ can be decomposed into the product of a Hermitian matrix and a unitary matrix, namely,
\begin{eqnarray}
V = ({\bf 1} - \eta) \cdot V^\prime \; ,
\label{eq:hermitianpara}
\end{eqnarray}
where $\eta$ is Hermitian and $V^\prime$ is unitary. Mathematically, this is just a direct consequence of the polar decomposition theorem. In the present case, the $3\times 3$ Hermitian matrix $\eta$ measures the strength of unitarity violation, and thus has been strictly constrained by current experimental observations, as we shall explain later on.

On the other hand, it has been proposed that the non-unitary mixing matrix $V$ can also be decomposed as below~\cite{Escrihuela:2015wra, Fernandez-Martinez:2016lgt}
\begin{eqnarray}
V = \left( \begin{matrix} \alpha^{}_{11} & 0 & 0 \cr \alpha^{}_{21} & \alpha^{}_{22} & 0 \cr \alpha^{}_{31} & \alpha^{}_{32} & \alpha^{}_{33} \end{matrix} \right) \cdot \widetilde{V} \equiv T \cdot \widetilde{V} \; ,
\label{eq:trianglepara}
\end{eqnarray}
where $T$ is by definition a lower-triangular matrix and $\widetilde{V}$ is a unitary matrix. Notice that the matrix elements of $T$ have been explicitly given in Eq.~(\ref{eq:trianglepara}), where $\alpha^{}_{ji} = 0$ (for $1 \leq j < i \leq 3$), $\alpha^{}_{ii}$ (for $i = 1, 2, 3$) are real and positive numbers, whereas $\alpha^{}_{ji}$ (for $1 \leq i < j \leq 3$) are complex numbers. In fact, the decomposition in Eq.~(\ref{eq:trianglepara}) is known as the QR factorization of an arbitrary $3\times 3$ complex matrix. It is worthwhile to stress that the unitary matrix $V^\prime$ in Eq.~(\ref{eq:hermitianpara}) and $\widetilde{V}$ in Eq.~(\ref{eq:trianglepara}) must be different, since the associated Hermitian matrix $({\bf 1} - \eta)$ and the lower-triangular matrix $T$ cannot be exactly identical.

An immediate question is then how the decompositions in Eq.~(\ref{eq:hermitianpara}) and Eq.~(\ref{eq:trianglepara}) are related to each other. Such a correspondence can be established by performing the QR factorization of the Hermitian matrix $({\bf 1} - \eta)$. Following the standard procedure, as summarized in Appendix~\ref{app:QR}, we can obtain
\begin{eqnarray}
{\bf 1} - \eta \approx \begin{pmatrix}
		1-\eta_{ee}^{} && 0 && 0 \\
		-2 \eta_{e\mu}^* && 1-\eta_{\mu\mu}^{} && 0 \\
		-2 \eta_{e\tau}^* && -2 \eta_{\mu\tau}^* && 1-\eta_{\tau\tau}^{}
	\end{pmatrix} \cdot \begin{pmatrix}
	1 && -\eta_{e\mu}^{} && -\eta_{e\tau}^{} \\
	+\eta_{e\mu}^\ast && 1 && -\eta_{\mu\tau}^{}\\
	+\eta_{e\tau}^\ast && +\eta_{\mu\tau}^\ast && 1
    \end{pmatrix} \;,
\label{eq:RQ}
\end{eqnarray}
where all the higher-order terms of ${\cal O}(|\eta^{}_{\alpha\beta}|^2)$ for $\alpha, \beta = e, \mu, \tau$ have been omitted in the lower-triangular matrix and the unitary matrix on the right-hand side. Substituting Eq.~(\ref{eq:RQ}) into Eq.~(\ref{eq:hermitianpara}) and comparing the latter with Eq.~(\ref{eq:trianglepara}), one can arrive at
\begin{eqnarray}
T \approx \begin{pmatrix}
		1-\eta_{ee}^{} && 0 && 0 \\
		-2 \eta_{e\mu}^* && 1-\eta_{\mu\mu}^{} && 0 \\
		-2 \eta_{e\tau}^* && -2 \eta_{\mu\tau}^* && 1-\eta_{\tau\tau}^{}
	\end{pmatrix} \; , \quad \widetilde{V} \approx \begin{pmatrix}
	1 && -\eta_{e\mu}^{} && -\eta_{e\tau}^{} \\
	+\eta_{e\mu}^\ast && 1 && -\eta_{\mu\tau}^{}\\
	+\eta_{e\tau}^\ast && +\eta_{\mu\tau}^\ast && 1
    \end{pmatrix} V^\prime \; . \quad
\label{eq:approxid}
\end{eqnarray}
Some comments on the relations in Eq.~(\ref{eq:approxid}) are in order. The triangular parametrization of $V$ in Eq.~(\ref{eq:trianglepara}) can be related to the Hermitian parametrization in Eq.~(\ref{eq:hermitianpara}) by identifying $\alpha^{}_{11} = 1 - \eta^{}_{ee}$, $\alpha^{}_{22} = 1 - \eta^{}_{\mu\mu}$ and $\alpha^{}_{33} = 1 - \eta^{}_{\tau\tau}$ for three diagonal elements, and $\alpha^{}_{21} = -2\eta^*_{e\mu}$, $\alpha^{}_{31} = -2\eta^*_{e\tau}$ and $\alpha^{}_{32} = -2\eta^*_{\mu\tau}$ for three nonzero off-diagonal elements. This identification has already been observed in Refs.~\cite{Fernandez-Martinez:2016lgt, Blennow:2016jkn}. However, it should be further noticed that the unitary matrices $V^\prime$ and $\widetilde{V}$ will differ by some corrections of ${\cal O}(|\eta^{}_{\alpha\beta}|)$~\cite{Blennow:2016jkn}. On this point, we make two helpful remarks.
\begin{itemize}
\item As shown in Ref.~\cite{Antusch:2006vwa, Antusch:2014woa, Escrihuela:2015wra, Fernandez-Martinez:2016lgt}, the most stringent constraints on the unitarity violation arise from the electroweak precision data on the lepton universality and lepton-flavor-violating decays of charged leptons. No matter which process is considered, the non-unitary mixing matrix will always be involved as the combination $\left(VV^\dagger\right)^{}_{\alpha \beta}$, implying that the unitary matrix $V^\prime$ in the Hermitian parametrization of $V$ in Eq.~(\ref{eq:hermitianpara}) will be cancelled out. This also happens for $\widetilde{V}$ in the triangular parametrization in Eq.~(\ref{eq:trianglepara}). Therefore, the experimental constraints on the unitarity violation in these two different parametrizations will be equivalent after the identification of $T$ in Eq.~(\ref{eq:approxid}) is taken into account.

\item Once neutrino flavor oscillations are considered, it will be in principle problematic to take the standard parametrization for both $V^\prime$ and $\widetilde{V}$ and identify the corresponding mixing angles and CP-violating phases\cite{Blennow:2016jkn}. As indicated in Eq.~(\ref{eq:approxid}), the difference between them is on the order of ${\cal O}(|\eta^{}_{\alpha \beta}|)$. Hence it is hopefully possible to distinguish between the mixing parameters in $V^\prime$ and those in $\widetilde{V}$ in future neutrino oscillation experiments. This is because the oscillation probabilities depend on the mixing matrix $V$ itself instead of the combination $VV^\dagger$. Nevertheless, the parameters $|\eta^{}_{\alpha \beta}|$ have been restricted by the electroweak precision data to be smaller than ${\cal O}(10^{-3})$, which are too small to be essentially observed in current neutrino oscillation experiments. After the latest experimental constraints on $\eta^{}_{\alpha \beta}$ are taken into account, it is rather safe to ignore their corrections to the unitary matrix $\widetilde{V}$ in the triangular parametrization.
\end{itemize}
Although the triangular parametrization of the non-unitary mixing matrix $V$ in Eq.~(\ref{eq:trianglepara}) is just a straightforward implication of the QR factorization, it has actually been obtained for the first time in Refs.~\cite{Xing:2007zj, Xing:2011ur}, where the full parametrization of the $6\times 6$ unitary matrix in Eq.~(\ref{eq:unitary}) is proposed and the triangular parametrization of the $3\times 3$ sub-matrix $V$ naturally emerges.

For later convenience, we shall adopt a specific parametrization of the non-unitary mixing matrix $V$ and explain current experimental constraints on the parameters of unitarity violation. First, the Hermitian parametrization in Eq.~(\ref{eq:hermitianpara}) as advocated in Ref.~\cite{Antusch:2006vwa} will be implemented, and the parameters characterizing the unitarity violation are three real numbers $\{\eta^{}_{ee}, \eta^{}_{\mu\mu}, \eta^{}_{\tau\tau}\}$ and three complex ones $\{\eta^{}_{e\mu}, \eta^{}_{e\tau}, \eta^{}_{\mu\tau}\}$. Furthermore, as mentioned in Sec.~\ref{sec:intro}, the unitarity condition $VV^\dagger = {\bf 1} - RR^\dagger$ holds in the type-I seesaw model, so we can observe that $\left(VV^\dagger\right)^{}_{\alpha \alpha} = 1 - (|R^{}_{\alpha 1}|^2 + |R^{}_{\alpha 2}|^2 + |R^{}_{\alpha 3}|^2) \leq 1$. On the other hand, we have $\left(VV^\dagger\right)^{}_{\alpha \alpha} = 1 - 2\eta^{}_{\alpha \alpha} + \left(|\eta^{}_{\alpha e}|^2 + |\eta^{}_{\alpha \mu}|^2 + |\eta^{}_{\alpha \tau}|^2\right)$. Thus $\eta^{}_{\alpha \alpha} > 0$ must be satisfied for $|\eta^{}_{\alpha \beta}| \ll 1$. The global-fit analysis of these parameters in light of the electroweak precision data has been performed in Ref.~\cite{Fernandez-Martinez:2016lgt} and the final bounds in the general seesaw model have been obtained at the $2\sigma$ level, viz.
\begin{eqnarray}
0 \leq \eta^{}_{ee} < 1.25\times 10^{-3} \;, \quad 0 \leq \eta^{}_{\mu \mu} < 2.21\times 10^{-4} \; , \quad 0 \leq \eta^{}_{\tau \tau} < 2.81\times 10^{-3} \; ;
\label{eq:bounddiag}
\end{eqnarray}
and
\begin{eqnarray}
|\eta^{}_{e\mu}| < 1.20\times 10^{-5} \; , \quad |\eta^{}_{e\tau}| < 1.35\times 10^{-3} \; , \quad |\eta^{}_{\mu\tau}| < 6.13\times 10^{-4} \; .
\label{eq:boundoff}
\end{eqnarray}
The second step is to translate the above bounds on $|\eta^{}_{\alpha \beta}|$ into those on $\alpha^{}_{ij}$ (for $1\leq j \leq i \leq 3$). This can be simply achieved by using the first equality in Eq.~(\ref{eq:approxid}). At the $2\sigma$ level, we have
\begin{eqnarray}
0.99875 < \alpha^{}_{11} \leq 1\; , \quad 0.99978 < \alpha^{}_{22} \leq 1 \; , \quad 0.99719 < \alpha^{}_{33} \leq 1 \; ;
\label{eq:alphadiag}
\end{eqnarray}
and
\begin{eqnarray}
|\alpha^{}_{21}| < 2.40\times 10^{-5} \; , \quad |\alpha^{}_{31}| < 2.70\times 10^{-3} \; , \quad |\alpha^{}_{32}| < 1.23\times 10^{-3} \; .
\label{eq:alphaoff}
\end{eqnarray}
The derived bounds on $|\alpha^{}_{ij}|$ are well consistent with those given in Ref.~\cite{Escrihuela:2016ube}. It is worth mentioning that we shall define $\alpha^{}_{ij} \equiv |\alpha^{}_{ij}| e^{{\rm i}\phi^{}_{ij}}$ for $1 \leq j < i \leq 3$ and these three phases $\{\phi^{}_{21}, \phi^{}_{31}, \phi^{}_{32}\}$ are completely unconstrained by the electroweak precision data and lepton-flavor-violating decays. In the following discussions, these phases will be taken to be free parameters. Finally, as mentioned before, it is reasonable to ignore the difference between $V^\prime$ and $\widetilde{V}$ in light of the current bounds on $|\eta^{}_{\alpha \beta}|$ in Eq.~(\ref{eq:boundoff}). Therefore, both $V^\prime$ and $\widetilde{V}$ can be identified with the mixing matrix in the unitary limit, and we choose the standard parametrization of the unitary matrix $\widetilde{V}$ as advocated by the Particle Data Group~\cite{Zyla:2020zbs}, i.e.,
\begin{eqnarray}
\widetilde{V} = \begin{pmatrix} c_{12}^{} c_{13}^{} && s_{12}^{} c_{13}^{} && s_{13}^{} {\rm e}^{- {\rm i}\delta} \\ -s_{12}^{} c_{23}^{} - c_{12}^{} s_{13}^{} s_{23}^{} {\rm e}^{{\rm i}\delta} && +c_{12}^{} c_{23}^{} - s_{12}^{} s_{13}^{} s_{23}^{} {\rm e}^{{\rm i}\delta} && c_{13}^{} s_{23}^{} \\ +s_{12}^{} s_{23}^{} - c_{12}^{} s_{13}^{} c_{23}^{} {\rm e}^{{\rm i}\delta} && -c_{12}^{} s_{23}^{}-s_{12}^{} s_{13}^{} c_{23}^{} {\rm e}^{{\rm i}\delta} && c_{13}^{} c_{23}^{}
\end{pmatrix} \cdot \begin{pmatrix} e^{{\rm i}\rho} && 0 && 0 \\ 0 && e^{{\rm i}\sigma} && 0 \\ 0 && 0 && 1 \end{pmatrix} \; ,
\label{eq:standardpara}
\end{eqnarray}
where $c^{}_{ij} \equiv \cos \theta^{}_{ij}$ and $s^{}_{ij} \equiv \sin \theta^{}_{ij}$ (for $ij = 12, 13, 23$) have been defined. Three neutrino mixing angles $\{\theta^{}_{12}, \theta^{}_{13}, \theta^{}_{23}\}$ and the Dirac-type CP-violating phase $\delta$, together with two neutrino mass-squared differences, can be extracted from the global-fit analysis of neutrino oscillation data~\cite{Esteban:2020cvm}, while two Majorana-type CP-violating phases $\{\rho, \sigma\} \in [0, \pi)$ are essentially free. For illustration, we take the best-fit values from the global-fit results, namely,
\begin{eqnarray}
\theta_{12}^{} &=& 33.4^\circ_{} \; , \qquad \delta = 195^\circ_{} \; ,\nonumber \\
\theta_{23}^{} &=& 49.0^\circ_{} \; , \qquad \Delta m_{21}^2 = +7.42 \times 10^{-5}_{}~{\rm eV^2_{}}\nonumber \\
\theta_{13}^{} &=& 8.57^\circ_{} \; , \qquad \Delta m_{31}^2 = +2.51 \times 10^{-3}_{}~{\rm eV^2_{}} \; ,
\label{eq:mixparaNO}
\end{eqnarray}
for the normal neutrino mass ordering (NO) with $m^{}_1 < m^{}_2 < m^{}_3$;
\begin{eqnarray}
\theta_{12}^{} &=& 33.5^\circ_{} \; , \qquad \delta = 286^\circ_{} \; ,\nonumber \\
\theta_{23}^{} &=& 49.3^\circ_{} \; , \qquad \Delta m_{21}^2 = +7.42 \times 10^{-5}_{}~{\rm eV^2_{}}\nonumber \\
\theta_{13}^{} &=& 8.61^\circ_{} \; , \qquad \Delta m_{32}^2 = -2.50 \times 10^{-3}_{}~{\rm eV^2_{}} \; ,
\label{eq:mixparaIO}
\end{eqnarray}
for the inverted neutrino mass ordering (IO) with $m^{}_3 < m^{}_1 < m^{}_2$. Note that the neutrino mass-squared differences have been defined as $\Delta m^2_{ji} \equiv m^2_j - m^2_i$ for $ji = 21, 31, 32$. In our numerical calculations in the next section, we shall use the allowed ranges of the non-unitary parameters in Eqs.~(\ref{eq:alphadiag}) and (\ref{eq:alphaoff}), and the best-fit values of the ordinary mixing parameters in Eq.~(\ref{eq:mixparaNO}) in the NO case or those in Eq.~(\ref{eq:mixparaIO}) in the IO case.

\section{CP Asymmetries}
\label{sec:CP}

\subsection{General Remarks}

\subsubsection{Neutrino-neutrino Oscillations}
Once three ordinary neutrinos mix with extra heavy fermions, the flavor mixing matrix appearing in the leptonic charged-current interaction will be non-unitary~\cite{Langacker:1988ur}. The phenomenology of non-unitary leptonic flavor mixing has been studied extensively in the literature~\cite{Czakon:2001em, Giunti:2004zf, Antusch:2006vwa, Xing:2008fg, Luo:2008vp, Altarelli:2008yr, Bhattacharya:2009nu, Dev:2009aw, Xing:2009ce}. Now that the flavor mixing matrix is non-unitary, it will be more convenient to introduce the neutrino flavor eigenstates $|\nu^{}_\alpha\rangle$ for $\alpha = e, \mu, \tau$, i.e.,
\begin{eqnarray}
|\nu_\alpha^{}\rangle = \frac{1}{\sqrt{\left(VV^\dagger_{}\right)_{\alpha\alpha}^{}}}\sum_i V_{\alpha i}^\ast |\nu_i\rangle \; ,
\label{eq:nualpha}
\end{eqnarray}
which have been properly normalized. Though neutrino mass eigenstates are orthonormal, namely, $\langle \nu^{}_j|\nu^{}_i\rangle = \delta^{}_{ij}$, the neutrino flavor eigenstates are not orthogonal in the sense that $\langle \nu^{}_\beta | \nu^{}_\alpha\rangle \neq 0$ holds for $\alpha \neq \beta$, but $\langle \nu^{}_\alpha| \nu^{}_\alpha \rangle = 1$ for $\alpha = e, \mu, \tau$. Following Refs.~\cite{Antusch:2006vwa, Xing:2007zj}, one can solve the Schr\"{o}dinger-like equation for the time-evolved neutrino flavor eigenstate $|\nu^{}_\alpha(t)\rangle$ and then calculate the neutrino-neutrino oscillation amplitudes $\langle \nu^{}_\beta|\nu^{}_\alpha(t)\rangle$ at $t \approx L$ with $L$ being the distance between the neutrino source and the detector. Therefore, the oscillation probabilities $P(\nu_\alpha^{} \to \nu_\beta^{}) \equiv |\langle \nu_\beta^{} | \nu_\alpha^{} (L)\rangle|^2_{}$ in the presence of a non-unitary mixing matrix are given by~\cite{Xing:2007zj}
\begin{eqnarray}
P\left(\nu_\alpha^{} \to \nu_\beta^{}\right) = \frac{\displaystyle \sum_{i=1}^3 \left|V^{}_{\alpha i}\right|^2 \left|V^{}_{\beta i}\right|^2 + 2\sum_{i<j} {\rm Re} \left[V_{\alpha i}^{} V_{\alpha j}^\ast V_{\beta i}^\ast V_{\beta j}^{}\right] \cos F_{ji}^{} + 2 \sum_{i<j} J_{\alpha\beta}^{ij} \sin F_{ji}^{}}{\left(VV^\dagger_{}\right)_{\alpha\alpha}^{} \left(VV^\dagger_{}\right)_{\beta\beta}^{}} \,,
\label{eq:pnunu}
\end{eqnarray}
where the Jarlskog-like rephasing invariants $J_{\alpha\beta}^{ij} \equiv {\rm Im}\left[V_{\alpha i}^{} V_{\alpha j}^\ast V_{\beta i}^\ast V_{\beta j}^{}\right]$ have been defined (for $i, j = 1, 2, 3$) in a similar way to those for a unitary flavor mixing matrix~\cite{Jarlskog:1985ht, Wu:1985ea} and $F^{}_{ji} \equiv \Delta m^2_{ji} L/(2E)$ for $ji = 21, 31, 32$ are the oscillation phases. It is worthwhile to notice that the sign in front of the last term in the numerator on the right-hand side of Eq.~(\ref{eq:pnunu}) is different from that in Ref.~\cite{Xing:2007zj}, where $\sin F^{}_{ij}$ for $ij = 12, 23, 13$ have been used. As is well known, for a unitary mixing matrix, the rephasing invariants $J_{\alpha\beta}^{ij}$ are all equal up to a minus sign, implying that there is one unique Jarlskog invariant, usually denoted as ${\cal J}$. In contrast, for a non-unitary mixing matrix $V$, the unitarity conditions are no longer applicable, but the identities $J_{\alpha\beta}^{ij} = - J_{\beta\alpha}^{ij} = - J_{\alpha\beta}^{ji} = J_{\beta\alpha}^{ji}$ together with $J_{\alpha\alpha}^{ij} = J_{\alpha\beta}^{ii} = 0$ hold for $\alpha, \beta = e, \mu, \tau$ and $i, j = 1, 2, 3$ according to the definition of $J_{\alpha\beta}^{ij}$. Hence it is easy to verify that there are totally nine independent Jarlskog-like rephasing invariants $J_{\alpha\beta}^{ij}$ in the non-unitary case.

By using the triangular parametrization of $V$ in Eq.~(\ref{eq:trianglepara}), together with $\widetilde{V}$ in Eq.~(\ref{eq:standardpara}), one can write down the explicit expressions of nine independent $J^{ij}_{\alpha\beta}$. However, the exact expressions are lengthy and too complicated to be useful. In consideration of $s^2_{13} \approx 0.022$ and $|\alpha^{}_{ji}| \ll 1$ (for $1\leq i < j \leq 3$), we can safely neglect the higher-order terms and derive the approximate analytical expressions of $J_{\alpha\beta}^{ij}$. By employing the Jarlskog invariant $\mathcal{J} \approx s_{12}^{} c_{12}^{} s_{13}^{} s_{23}^{} c_{23}^{} \sin \delta$, we find that there are only seven independent Jarlskog-like invariants. More explicitly, we have
\begin{eqnarray}
J_{e\mu}^{23} \approx J_{e\mu}^{31} \approx \alpha_{11}^2 \alpha_{22}^2 \mathcal{J} \; , \quad	J_{e\mu}^{12} \approx J_{e\mu}^{23} - \alpha_{11}^2 \left| \alpha_{21}^{} \right| \alpha_{22}^{} s_{12}^{} c_{12}^{} c_{23}^{} \sin \phi_{21}^{} \; ,
\label{eq:Jemu}
\end{eqnarray}
for $(\alpha, \beta) = (e, \mu)$,
\begin{eqnarray}
J_{\tau e}^{23} \approx J_{\tau e}^{31} \approx  \alpha_{11}^2 \alpha_{33}^2 \mathcal{J} \; , \quad J_{\tau e}^{12} \approx J_{\tau e}^{23} - \alpha_{11}^2 \left| \alpha_{31}^{} \right| \alpha_{33}^{} s_{12}^{} c_{12}^{} s_{23}^{} \sin \phi_{31}^{} \; ,
\label{eq:Jetau}
\end{eqnarray}
for $(\alpha, \beta) = (\tau, e)$, and
\begin{eqnarray}
J_{\mu\tau}^{12} &\approx& \alpha_{22}^2 \alpha_{33}^2 \mathcal{J} - \alpha_{22}^2 \left| \alpha_{31}^{} \right| \alpha_{33}^{} s_{12}^{} c_{12}^{} s_{23}^{} c_{23}^2 \sin \phi_{31}^{} \; , \nonumber \\
J_{\mu\tau}^{23} &\approx& J_{\mu\tau}^{12} - \alpha_{22}^2 \left| \alpha_{32}^{} \right| \alpha_{33}^{} c_{12}^2 s_{23}^{} c_{23}^{} \sin \phi_{32}^{} \; , \quad
J_{\mu\tau}^{31} \approx J_{\mu\tau}^{12} + \alpha_{22}^2 \left| \alpha_{32}^{} \right| \alpha_{33}^{} s_{12}^2 s_{23}^{} c_{23}^{} \sin \phi_{32}^{} \; ,
\label{eq:Jmutau}
\end{eqnarray}
for $(\alpha, \beta) = (\mu, \tau)$. With the above Jarlskog-like invariants, one can compute the CP asymmetries for the probabilities of neutrino-neutrino $\nu^{}_\alpha \to \nu^{}_\beta$ oscillations and antineutrino-antineutrino $\overline{\nu}^{}_\alpha \to \overline{\nu}^{}_\beta$ oscillations. With a trivial CP-violating phase $\delta = 0$ or $\pi$, we have ${\cal J} = 0$ as in the unitary case with CP conservation. In the non-unitary case, although we have ${\cal J}^{23}_{e\mu} \approx {\cal J}^{31}_{e\mu} \approx {\cal J}^{23}_{\tau e} \approx {\cal J}^{31}_{\tau e} \approx 0$, CP violation is still present due to other non-vanishing Jarlskog-like parameters in Eqs.~(\ref{eq:Jemu})-(\ref{eq:Jmutau}). Further discussions about how to probe the non-unitarity induced CP violation in neutrino oscillation experiments can be found in Refs.~\cite{FernandezMartinez:2007ms, Antusch:2009pm, Blennow:2016jkn, Escrihuela:2016ube, Ellis:2020hus, Forero:2021azc}.

\subsubsection{Neutrino-antineutrino Oscillations}

As for the neutrino-antineutrino oscillations, there exist already extensive discussions in Refs.~\cite{Xing:2013ty,Schechter:1980gk,Bahcall:1978jn} in the case of a unitary mixing matrix. In the non-unitary case, the amplitudes need to be modified as
\begin{eqnarray}
A \left(\nu_\alpha^{} \to \overline{\nu}_\beta^{}\right) = \frac{K}{\sqrt{\left(VV^\dagger_{}\right)_{\alpha\alpha}^{} \left(VV^\dagger_{}\right)_{\beta\beta}^{}}} \sum_{i=1}^{3} V_{\alpha i}^\ast V_{\beta i}^\ast \frac{m^{}_i}{E} {\rm exp} \left(-{\rm i} \frac{m^2_i L}{2E} \right) \,,
    \label{eq:amp1}\\
A \left(\overline{\nu}_\alpha^{} \to \nu_\beta^{}\right) =  \frac{\overline{K}}{\sqrt{\left(VV^\dagger_{}\right)_{\alpha\alpha}^{} \left(VV^\dagger_{}\right)_{\beta\beta}^{}}} \sum_{i=1}^{3} V_{\alpha i}^{} V_{\beta i}^{} \frac{m^{}_i}{E} {\rm exp} \left(-{\rm i} \frac{m^2_i L}{2E} \right) \,,
	\label{eq:amp2}
\end{eqnarray}
where $K$ and $\overline K$ are the kinematical factors satisfying the identity $\left|K\right| = \left|\overline{K} \right|$. The oscillation amplitudes are obviously suppressed by the small ratios $m^{}_i/E$ (for $i = 1, 2, 3$) due to the helicity mismatches between neutrinos and antineutrinos. Therefore, given the amplitudes in Eqs.~(\ref{eq:amp1}) and (\ref{eq:amp2}), the probabilities of neutrino-antineutrino oscillations and their CP-conjugate counterparts are found to be
\begin{eqnarray}
P\left(\nu_\alpha^{} \to \overline{\nu}_\beta^{}\right) &=& \frac {\left| K \right|^2_{} }{\left( V V^\dagger_{} \right)_{\alpha\alpha}^{}\left(VV^\dagger_{}\right)_{\beta\beta}^{}} \left[ \frac{\left|\langle m \rangle_{\alpha\beta}^{} \right|^2_{}}{E^2} - 4 \sum_{i<j} \frac{m_i^{} m_j^{}}{E^2} \mathcal{C}_{\alpha \beta}^{ij} \sin^2_{} \frac{F_{ji}^{}}{2} + 2 \sum_{i<j}^3 \frac{m_i^{} m_j^{}}{E^2} \mathcal{V}_{\alpha\beta}^{ij}  \sin F_{ji}^{} \right] \; , \nonumber \\
&~& \\
P\left(\overline{\nu}_\alpha^{} \to \nu_\beta^{}\right) &=& \frac {\left| \overline{K} \right|^2_{} }{\left( V V^\dagger_{} \right)_{\alpha\alpha}^{}\left(VV^\dagger_{}\right)_{\beta\beta}^{}} \left[ \frac{\left|\langle m \rangle_{\alpha\beta}^{} \right|^2_{}}{E^2} - 4 \sum_{i<j} \frac{m_i^{} m_j^{}}{E^2} \mathcal{C}_{\alpha \beta}^{ij} \sin^2_{} \frac{F_{ji}^{}}{2} - 2 \sum_{i<j}^3 \frac{m_i^{} m_j^{}}{E^2} \mathcal{V}_{\alpha\beta}^{ij}  \sin F_{ji}^{} \right] \; , \nonumber \\
\end{eqnarray}
where $\mathcal{C}_{\alpha\beta}^{ij} \equiv {\rm Re}\left[V_{\alpha i}^{} V_{\beta i}^{} V_{\alpha j}^\ast V_{\beta j}^\ast\right]$ and $\mathcal{V}_{\alpha\beta}^{ij} \equiv {\rm Im}\left[V_{\alpha i}^{} V_{\beta i}^{} V_{\alpha j}^\ast V_{\beta j}^\ast\right]$ have been introduced, and the effective neutrino masses $\langle m \rangle_{\alpha\beta}^{} \equiv V_{\alpha 1}^{} V_{\beta 1}^{} m_1^{} + V_{\alpha 2}^{} V_{\beta 2}^{} m_2^{} + V_{\alpha 3}^{} V_{\beta 3}^{} m_3^{}$ (for $\alpha, \beta = e, \mu, \tau$) have been defined. Then, the CP asymmetries for neutrino-antineutrino oscillations turn out to be
\begin{eqnarray}
\mathcal{A}_{\alpha \beta}^{} \equiv \frac{P\left(\nu_\alpha^{} \to \overline{\nu}_\beta^{}\right) - P\left(\overline{\nu}_\alpha^{} \to \nu_\beta^{}\right)}{P\left(\nu_\alpha^{} \to \overline{\nu}_\beta^{}\right) + P\left(\overline{\nu}_\alpha^{} \to \nu_\beta^{}\right)} = \frac{\displaystyle 2\sum_{i<j} m_i^{} m_j^{} \mathcal{V}_{\alpha \beta}^{ij} \sin F_{ji}^{}}{\displaystyle \left|\langle m \rangle_{\alpha\beta}^{}\right|^2_{} - 4 \sum_{i<j} m_i^{} m_j^{} \mathcal{C}_{\alpha \beta}^{ij} \sin^2_{} \frac{F_{ji}^{}}{2}} \; ,
\label{ACPnonuni}
\end{eqnarray}	
where the normalization factors $(VV^\dagger)_{\alpha \alpha} (VV^\dagger)_{\beta\beta}$ due to the non-unitarity of the mixing matrix are cancelled out. It is worthwhile to mention that the CP asymmetries ${\cal A}^{}_{\alpha \beta}$ in Eq.~(\ref{ACPnonuni}) are formally the same as those derived in Refs.~\cite{Xing:2013ty, Xing:2013woa} in the unitary case, but with a non-unitary flavor mixing matrix $V$ involved in the parameters $\mathcal{C}_{\alpha\beta}^{ij} \equiv {\rm Re}\left[V_{\alpha i}^{} V_{\beta i}^{} V_{\alpha j}^\ast V_{\beta j}^\ast\right]$ and $\mathcal{V}_{\alpha\beta}^{ij} \equiv {\rm Im}\left[V_{\alpha i}^{} V_{\beta i}^{} V_{\alpha j}^\ast V_{\beta j}^\ast\right]$. Furthermore, although the normalization factors $(VV^\dagger)_{\alpha \alpha} (VV^\dagger)_{\beta\beta}$ are cancelled out in the CP asymmetries, they {\it do} appear in the oscillation probabilities. For instance, they will lead to the corrections to the zero-distance effects for $L = 0$, namely,
\begin{eqnarray}
P(\nu^{}_\alpha \to \overline{\nu}^{}_\beta) = P(\overline{\nu}^{}_\alpha \to \nu^{}_\beta) = \frac {\left| K \right|^2_{} }{\left( V V^\dagger_{} \right)_{\alpha\alpha}^{}\left(VV^\dagger_{}\right)_{\beta\beta}^{}}  \frac{\left|\langle m \rangle_{\alpha\beta}^{} \right|^2_{}}{E^2} \; .
\end{eqnarray}
Hence the zero-distance effects receive also the contributions from the non-unitarity of the flavor mixing matrix, which has been observed in the case of neutrino-neutrino oscillations~\cite{Antusch:2006vwa}.

Since the CP asymmetries ${\cal A}^{}_{\alpha \beta}$ in Eq.~(\ref{ACPnonuni}) depend crucially on the Jarlskog-like parameters $\mathcal{V}_{\alpha \beta}^{ij}$, it is necessary to examine the basic properties of the latter. First of all, according to their definitions, the CP-conserving parameters $\mathcal{C}_{\alpha \beta}^{ij}$ and the CP-violating Jarlskog-like parameters $\mathcal{V}_{\alpha \beta}^{ij}$ fulfill the following identities
\begin{eqnarray}
\mathcal{C}_{\alpha \beta}^{ij} &=& \mathcal{C}_{\beta \alpha}^{ij} = +\mathcal{C}_{\alpha \beta}^{ji} = +\mathcal{C}_{\beta \alpha}^{ji}\,,\\
	\mathcal{V}_{\alpha \beta}^{ij} &=& \mathcal{V}_{\beta \alpha}^{ij} = -\mathcal{V}_{\alpha \beta}^{ji} = -\mathcal{V}_{\beta \alpha}^{ji}\, ,
\end{eqnarray}
where $\alpha, \beta = e, \mu, \tau$ and $i, j = 1, 2, 3$. Without the unitarity constraints from the mixing matrix $V$, one obtains totally 18 independent parameters $\mathcal{V}_{\alpha \beta}^{ij}$, which can be regarded as the $(\alpha, \beta)$-elements of three real and symmetric $3\times 3$ matrices $\mathcal{V}_{}^{12}$, $\mathcal{V}_{}^{13}$ and $\mathcal{V}_{}^{23}$.

To derive the explicit expressions of ${\cal V}^{ij}_{\alpha \beta}$, one can insert the parametrization of the non-unitary mixing matrix $V$ in Eq.~(\ref{eq:trianglepara}) and Eq.~(\ref{eq:standardpara}) into their definitions. Then, one can further subtract the results in the unitary limit, which have already been calculated in Refs.~\cite{Xing:2013ty, Xing:2013woa} and collected in Appendix~\ref{app:UN} for reference. More explicitly, we introduce the differences between the expressions in the non-unitary case and those in the unitary case, namely,
\begin{eqnarray}
\epsilon_{\alpha \beta}^{ij} \equiv \mathcal{V}_{\alpha \beta}^{ij} - \widetilde{\mathcal{V}}_{\alpha \beta}^{ij} \; ,
\end{eqnarray}
where $\widetilde{\mathcal{V}}_{\alpha \beta}^{ij}$ refer to the Jarlskog-like parameters in the unitary case. Notice that only nine out of 18 parameters $\widetilde{\mathcal{V}}_{\alpha \beta}^{ij}$ are independent because of the unitarity of the flavor mixing matrix~~\cite{Xing:2013ty, Xing:2013woa}, which can be chosen to be $\widetilde{\cal V}^{ij}_{\alpha \alpha}$ (for $ij = 12, 13, 23$ and $\alpha = e, \mu, \tau$) and their expressions in terms of mixing parameters are given in Appendix~\ref{app:UN}. In consideration of current experimental constraints on the unitarity-violating parameters summarized in Sec.~\ref{sec:para} and $s^2_{13} \sim 10^{-2}$, we retain only the leading-order terms in $\epsilon^{ij}_{\alpha \beta}$, i.e.,
\begin{eqnarray}
	\epsilon_{ee}^{12} &=& \left(\alpha_{11}^4-1\right) s_{12}^2 c_{12}^2 c_{13}^4 \sin 2(\rho-\sigma)\,, \label{eq:eee12}\quad\\
	\epsilon_{ee}^{13} &=& \left(\alpha_{11}^4-1\right) c_{12}^2 s_{13}^2 c_{13}^2 \sin 2(\rho+\delta)\,, \label{eq:eee13}\quad\\
	\epsilon_{ee}^{23} &=& \left(\alpha_{11}^4-1\right) s_{12}^2 s_{13}^2 c_{13}^2 \sin 2(\sigma+\delta)\, , \label{eq:eee23}
\end{eqnarray}	
for $(\alpha, \beta) = (e, e)$;
\begin{eqnarray}	
	\epsilon_{\mu\mu}^{12} &\approx& \left(\alpha_{22}^4-1\right) c_{23}^2\left[ s_{12}^2 c_{12}^2 c_{23}^2 \sin 2(\rho-\sigma) + 2 \mathcal{J}_{12}^{} \right]  - 2\left| \alpha_{21}^{} \right| \alpha_{22}^3 s_{12}^{} c_{12}^{} c_{23}^3 \Phi_{21}^{} \,, \quad
	\label{eq:emm12}\\
	\epsilon_{\mu\mu}^{13} &\approx& \left(\alpha_{22}^4-1\right) s_{23}^2 \left[s_{12}^2 c_{23}^2 \sin 2\rho + 2\mathcal{J}^{}_{\rm r}\sin(2\rho + \delta)\right] - 2 \left| \alpha_{21}^{} \right| \alpha_{22}^3 s_{12}^{} c_{12}^{} s_{23}^2 c_{23}^{} \sin \left(2\rho + \phi_{21}^{}\right)\,, \quad \quad
	\label{eq:emm13}\\
	\epsilon_{\mu\mu}^{23} &\approx& \left(\alpha_{22}^4-1\right)  s_{23}^2 \left[c_{12}^2 c_{23}^2 \sin 2\sigma - 2\mathcal{J}^{}_{\rm r} \sin(2\sigma + \delta)\right] + 2 \left| \alpha_{21}^{} \right| \alpha_{22}^3 s_{12}^{} c_{12}^{} s_{23}^2 c_{23}^{} \sin \left(2\sigma + \phi_{21}^{}\right)\,, \quad \quad
	\label{eq:emm23}
\end{eqnarray}
for $(\alpha, \beta) = (\mu, \mu)$;
\begin{eqnarray}
	\epsilon_{\tau\tau}^{12} &\approx& \left(\alpha_{33}^4-1\right) s_{23}^2 \left[ s_{12}^2 c_{12}^2 s_{23}^2 \sin 2(\rho - \sigma) - 2\mathcal{J}_{12}^{} \right]  + 2 \left| \alpha_{31}^{} \right| \alpha_{33}^3 s_{12}^{} c_{12}^{} s_{23}^3 \Phi_{31}^{} \nonumber\\
	&& - 4 \left| \alpha_{32}^{} \right| \alpha_{33}^3 s_{12}^2 c_{12}^2 s_{23}^3 c_{23}^{} \sin 2(\rho-\sigma) \cos \phi_{32}^{} \,, \quad \quad \label{eq:ett12}\\
	\epsilon_{\tau\tau}^{13} &\approx& \left(\alpha_{33}^4-1\right) c_{23}^2 \left[s_{12}^2 s_{23}^2 \sin2\rho - 2\mathcal{J}^{}_{\rm r} \sin(2\rho +\delta)\right]  + 2 \left| \alpha_{31}^{} \right| \alpha_{33}^3 s_{12}^{} c_{12}^{} s_{23}^{} c_{23}^2 \sin\left( 2\rho + \phi_{31}^{} \right) \quad \quad \nonumber\\
	&&+ 2\left| \alpha_{32}^{} \right| \alpha_{33}^3 s_{12}^2 s_{23}^{} c_{23}^{}\left[s_{23}^2 \sin \left(2\rho -\phi_{32}^{} \right) - c_{23}^2 \sin \left(2\rho +\phi_{32}^{} \right)\right]\,, \label{eq:ett13} \\
	\epsilon_{\tau\tau}^{23} &\approx& \left(\alpha_{33}^4-1\right) c_{23}^2 \left[c_{12}^2 s_{23}^2 \sin2\sigma + 2\mathcal{J}^{}_{\rm r} \sin(2\sigma +\delta)\right]  - 2 \left| \alpha_{31}^{} \right| \alpha_{33}^3 s_{12}^{} c_{12}^{} s_{23}^{} c_{23}^2 \sin\left( 2\sigma + \phi_{31}^{} \right)\quad \quad \nonumber\\
	&& + 2\left| \alpha_{32}^{} \right| \alpha_{33}^3 c_{12}^2 s_{23}^{} c_{23}^{} \left[s_{23}^2 \sin \left(2\sigma -\phi_{32}^{} \right) - c_{23}^2 \sin \left(2\sigma +\phi_{32}^{} \right)\right]\,,
	\label{eq:ett23}
\end{eqnarray}	
for $(\alpha, \beta) = (\tau, \tau)$. In the above formulas, we have introduced the reduced Jarlskog invariant ${\cal J}^{}_{\rm r} \equiv {\cal J}/\sin\delta \approx s^{}_{12} c^{}_{12} s^{}_{13} s^{}_{23} c^{}_{23}$ and \begin{eqnarray}
	\mathcal{J}_{12}^{} &\equiv& \mathcal{J}^{}_{\rm r} \left[ c_{12}^2 \sin \left( 2\rho - 2\sigma + \delta \right) - s_{12}^2 \sin \left( 2\rho - 2\sigma - \delta \right) \right] \,, \label{eq:J12}\\
	\Phi_{21}^{} &\equiv& c_{12}^2 \sin \left( 2\rho - 2\sigma + \phi_{21}^{} \right) - s_{12}^2 \sin \left( 2\rho - 2\sigma - \phi_{21}^{} \right) \,,\\
	\Phi_{31}^{} &\equiv& c_{12}^2 \sin \left( 2\rho - 2\sigma + \phi_{31}^{} \right) - s_{12}^2 \sin \left( 2\rho - 2\sigma - \phi_{31}^{} \right) \,.
\end{eqnarray}
As shown in Appendix~\ref{app:UN}, the nine off-diagonal parameters $\widetilde{\cal V}^{12}_{\alpha \beta}$, $\widetilde{\cal V}^{13}_{\alpha \beta}$ and $\widetilde{\cal V}^{23}_{\alpha \beta}$ for $(\alpha, \beta) = (e, \mu)$, $(e, \tau)$ and $(\mu, \tau)$ are not independent but related to $\widetilde{\cal V}^{12}_{\alpha \alpha}$, $\widetilde{\cal V}^{13}_{\alpha \alpha}$ and $\widetilde{\cal V}^{23}_{\alpha \alpha}$ for $\alpha = e, \mu, \tau$. In contrast, their counterparts in the non-unitary case are actually independent. In the leading-order approximation, we obtain
\begin{eqnarray}
	\epsilon_{e\mu}^{12} &\approx& \left(1-\alpha_{11}^2 \alpha_{22}^2\right) c_{13}^2 \left[s_{12}^2 c_{12}^2 \left(c_{23}^2 - s_{13}^2 s_{23}^2\right) \sin 2(\rho-\sigma) + \mathcal{J}_{12}^{} \right] + \alpha_{11}^2 \left| \alpha_{21}^{} \right| \alpha_{22}^{} s_{12}^{} c_{12}^{} c_{23}^{} \Phi_{21}^{} \,, \quad \quad \label{eq:eem12}\\
	\epsilon_{e\mu}^{13} &\approx& \left(1 - \alpha_{11}^2 \alpha_{22}^2\right) c_{13}^2 \left[\mathcal{J}^{}_{\rm r} \sin (2\rho + \delta) - c_{12}^2 s_{13}^2 s_{23}^2 \sin 2\left( \rho + \delta \right) \right]\,, \label{eq:eem13}\quad\\
	\epsilon_{e\mu}^{23} &\approx& \left(\alpha_{11}^2 \alpha_{22}^2 - 1\right) c_{13}^2  \left[\mathcal{J}^{}_{\rm r} \sin (2\sigma + \delta) - s_{12}^2 s_{13}^2 s_{23}^2 \sin 2\left( \sigma + \delta \right) \right]\,,
	\label{eq:eem23}
\end{eqnarray}
for $(\alpha, \beta) = (e, \mu)$;
\begin{eqnarray}
\epsilon_{e\tau}^{12} &\approx& \left(1 - \alpha_{11}^2 \alpha_{33}^2\right) \left[ s_{12}^2 c_{12}^2 s_{23}^2 \sin 2(\rho-\sigma) - \mathcal{J}_{12}^{} \right]  - \alpha_{11}^2 \left| \alpha_{31}^{} \right| \alpha_{33}^{} s_{12}^{} c_{12}^{} s_{23}^{} \Phi_{31}^{} \nonumber\\
	&& + 2\alpha_{11}^2 \left| \alpha_{32}^{} \right| \alpha_{33}^{} s_{12}^2 c_{12}^2 s_{23}^{} c_{23}^{} \sin 2(\rho - \sigma) \cos \phi_{32}^{}\,, \label{eq:eet12}\\
\epsilon_{e\tau}^{13} &\approx& \left(\alpha_{11}^2 \alpha_{33}^2 - 1\right) \mathcal{J}^{}_{\rm r} \sin (2\rho + \delta) + \alpha_{11}^2 \alpha_{33}^{} c_{12}^{} s_{13}^{}  \left\{\left| \alpha_{31}^{} \right| c_{12}^{} c_{23}^{} \sin \left( 2\rho + \delta + \phi_{31}^{} \right) \right. \nonumber \\
&~& \left. - \left| \alpha_{32}^{} \right| s_{12}^{} \left[ c_{23}^2 \sin \left( 2\rho + \delta + \phi_{32}^{} \right) - s_{23}^2 \sin \left( 2\rho + \delta - \phi_{32}^{} \right) \right] \right\} \,, \label{eq:eet13} \\
\epsilon_{e\tau}^{23} &\approx& \left(1 - \alpha_{11}^2 \alpha_{33}^2\right) \mathcal{J}^{}_{\rm r} \sin (2\sigma + \delta) + \alpha_{11}^2 \alpha_{33}^{} s_{12}^{} s_{13}^{} \left\{ \left| \alpha_{31}^{} \right| s_{12}^{} c_{23}^{} \sin\left(2\sigma+\delta+\phi_{31}^{}\right) \right. \nonumber \\
&~& \left. + \left| \alpha_{32}^{} \right| c_{12}^{} \left[ c_{23}^2 \sin \left( 2\sigma + \delta + \phi_{32}^{} \right) - s_{23}^2 \sin \left( 2\sigma + \delta - \phi_{32}^{} \right) \right] \right\}\,, \label{eq:eet23}
\end{eqnarray}
for $(\alpha, \beta) = (e, \tau)$;
\begin{eqnarray}
\epsilon_{\mu\tau}^{12} &\approx& \left(\alpha_{22}^2 \alpha_{33}^2 - 1\right) \left[s_{12}^2 c_{12}^2 s_{23}^2 c_{23}^2 \sin 2(\rho-\sigma) - \mathcal{J}_{12}^{} \cos 2 \theta_{23}^{}  \right]  + \alpha_{22}^2 \left| \alpha_{31}^{} \right| \alpha_{33}^{} s_{12}^{} c_{12}^{} s_{23}^{} c_{23}^2 \Phi_{31}^{} \nonumber\\
&& - 2\alpha_{22}^2 \left| \alpha_{32}^{} \right| \alpha_{33}^{} s_{12}^2 c_{12}^2 s_{23}^{} c_{23}^3 \sin 2(\rho - \sigma) \cos \phi_{32}^{}\,, \label{eq:emt12} \\
\epsilon_{\mu\tau}^{13} &\approx& \left(1 - \alpha_{22}^2 \alpha_{33}^2\right) \left[ s_{12}^2 s_{23}^2 c_{23}^2 \sin2\rho -  \mathcal{J}^{}_{\rm r} \cos 2\theta_{23}^{} \sin (2\rho + \delta) \right] \nonumber \\
&~& - \alpha_{22}^2  \left| \alpha_{31}^{} \right| \alpha_{33}^{} s_{12}^{} c_{12}^{} s_{23}^{} c_{23}^2 \sin\left(2\rho + \phi_{31}^{}\right) \nonumber\\
&& + \alpha_{22}^2  \left| \alpha_{32}^{} \right| \alpha_{33}^{} s_{12}^2 s_{23}^{} c_{23}^{} \left[c_{23}^2 \sin\left(2\rho + \phi_{32}^{}\right) - s_{23}^2 \sin \left(2\rho - \phi_{32}^{}\right)\right]\,, \label{eq:emt13}\\
\epsilon_{\mu\tau}^{23} &\approx& \left(1 - \alpha_{22}^2 \alpha_{33}^2\right) \left[ c_{12}^2 s_{23}^2 c_{23}^2 \sin 2\sigma + \mathcal{J}^{}_{\rm r} \cos 2\theta_{23}^{} \sin (2\sigma + \delta) \right] \nonumber \\
&~&+ \alpha_{22}^2  \left| \alpha_{31}^{} \right| \alpha_{33}^{} s_{12}^{} c_{12}^{} s_{23}^{} c_{23}^2 \sin\left(2\sigma + \phi_{31}^{}\right) \nonumber\\
&& + \alpha_{22}^2  \left| \alpha_{32}^{} \right| \alpha_{33}^{} c_{12}^2 s_{23}^{} c_{23}^{} \left[c_{23}^2 \sin\left(2\sigma + \phi_{32}^{}\right) - s_{23}^2 \sin \left(2\sigma - \phi_{32}^{}\right)\right] \,,
    \label{eq:emt23}
\end{eqnarray}
for $(\alpha, \beta) = (\mu, \tau)$. Some comments on the approximate formulas of $\epsilon^{ij}_{\alpha \beta}$ are helpful.
\begin{itemize}
\item In the absence of unitarity violation, i.e., $\alpha^{}_{ii} = 1$ (for $i = 1, 2, 3$) and $|\alpha^{}_{ij}| = 0$ (for $ij = 21, 31, 32$), one can immediately verify that all the parameters $\epsilon^{ij}_{\alpha \beta}$ vanish as they should. In this case, the Jarlskog-like parameters ${\cal V}^{ij}_{\alpha \beta}$ are reduced to $\widetilde{\cal V}^{ij}_{\alpha \beta}$ in the unitary limit.
\item As far as the CP-violating phases $\{\rho, \sigma, \delta\}$ are concerned, we observe that only the phase difference $\rho - \sigma$ and the phase $\delta$ are involved in $\mathcal{V}_{\alpha \beta}^{12}$. In particular, $\mathcal{V}_{ee}^{12}$ depends only on the phase difference $\rho - \sigma$. Moreover, $\mathcal{V}_{\alpha \beta}^{13}$ contain the CP-violating phases $\rho$ and $\delta$, whereas $\mathcal{V}_{\alpha \beta}^{23}$ are dependent on $\sigma$ and $\delta$. On the other hand, if we focus on the extra CP-violating phases $\{\phi^{}_{21}, \phi^{}_{31}, \phi^{}_{32}\}$ from unitarity violation, the parametrization in Eq.~(\ref{eq:trianglepara}) indicates that the non-unitary phase $\phi_{21}^{}$ is only involved in $\mathcal{V}_{e \mu}^{ij}$, $\mathcal{V}_{\mu \mu}^{ij}$ and $\mathcal{V}_{\mu \tau}^{ij}$, while $\phi_{31}^{}$ and $\phi_{32}^{}$ are only contained in $\mathcal{V}_{e \tau}^{ij}$, $\mathcal{V}_{\tau \tau}^{ij}$ and $\mathcal{V}_{\mu \tau}^{ij}$.

\item A particularly interesting scenario is to assume all the ordinary CP-violating phases to be trivial, i.e., $\rho = \sigma = 0$ (or $\pi/2$) and $\delta = 0$ (or $\pi$). In this case, the Jarlskog-like parameters $\widetilde{\cal V}^{ij}_{\alpha \beta}$ in the unitary limit are all vanishing, so we have ${\cal V}^{ij}_{\alpha \beta} = \epsilon^{ij}_{\alpha \beta}$. With the help of the approximate formulas for $\epsilon^{ij}_{\alpha \beta}$, we can get
\begin{eqnarray}
\mathcal{V}_{ee}^{12} \approx \mathcal{V}_{ee}^{13} \approx \mathcal{V}_{ee}^{23} \approx \mathcal{V}_{e\mu}^{13} \approx \mathcal{V}_{e\mu}^{23} \approx 0 \label{eq:Vee230} \; .
\end{eqnarray}
In addition, the other non-vanishing Jarlskog-like parameters are given by
\begin{eqnarray}
\mathcal{V}_{\mu\mu}^{12} &\approx& -2\left| \alpha_{21}^{} \right| \alpha_{22}^3 s_{12}^{} c_{12}^{} c_{23}^3 \sin\phi_{21}^{} \, , \\
\mathcal{V}_{\mu\mu}^{13} &\approx& -\mathcal{V}_{\mu\mu}^{23} \approx  -2\left| \alpha_{21}^{} \right| \alpha_{22}^3 s_{12}^{} c_{12}^{} s_{23}^2 c_{23}^{} \sin\phi_{21}^{}, \label{eq:Vmm0}\\
\mathcal{V}_{\tau\tau}^{12} &\approx& +2\left| \alpha_{31}^{} \right| \alpha_{33}^3 s_{12}^{} c_{12}^{} s_{23}^3 \sin\phi_{31}^{}\,,\\
\mathcal{V}_{\tau\tau}^{13} &\approx& +2\left| \alpha_{31}^{} \right| \alpha_{33}^3 s_{12}^{} c_{12}^{} s_{23}^{} c_{23}^2 \sin\phi_{31}^{} -
		2\left| \alpha_{32}^{} \right| \alpha_{33}^3 s_{12}^2 s_{23}^{} c_{23}^{} \sin\phi_{32}^{}\,,\\
\mathcal{V}_{\tau\tau}^{23} &\approx& -2\left| \alpha_{31}^{} \right| \alpha_{33}^3 s_{12}^{} c_{12}^{} s_{23}^{} c_{23}^2 \sin\phi_{31}^{} -
		2\left| \alpha_{32}^{} \right| \alpha_{33}^3 c_{12}^2 s_{23}^{} c_{23}^{} \sin\phi_{32}^{}\,,\label{eq:tilVtt230}\\
\mathcal{V}_{e\mu}^{12} &\approx& +\alpha_{11}^2 \left| \alpha_{21}^{} \right| \alpha_{22}^{} s_{12}^{} c_{12}^{} c_{23}^{} \sin\phi_{21}^{}\,, \\
\mathcal{V}_{e\tau}^{12} &\approx& -\alpha_{11}^2 \left| \alpha_{31}^{} \right| \alpha_{33}^{} s_{12}^{} c_{12}^{} s_{23}^{} \sin\phi_{31}^{}\,,\\
		\mathcal{V}_{e\tau}^{13} &\approx& +\alpha_{11}^2 \alpha_{33}^{} c_{12}^{} s_{13}^{} \left(\left|\alpha_{31}^{}\right| c_{12}^{} c_{23}^{} \sin \phi_{31}^{} - \left|\alpha_{32}^{}\right| s_{12}^{} \sin \phi_{32}^{} \right)\\
		\mathcal{V}_{e\tau}^{23} &\approx& +\alpha_{11}^2 \alpha_{33}^{} s_{12}^{} s_{13}^{} \left(\left|\alpha_{31}^{}\right| s_{12}^{} c_{23}^{} \sin \phi_{31}^{} + \left|\alpha_{32}^{}\right| c_{12}^{} \sin \phi_{32}^{} \right)\\
		\mathcal{V}_{\mu\tau}^{12} &\approx& +\alpha_{22}^2 \left| \alpha_{31}^{} \right| \alpha_{33}^{} s_{12}^{} c_{12}^{} s_{23}^{} c_{23}^2 \sin\phi_{31}^{}\,,\\
		\mathcal{V}_{\mu\tau}^{13} &\approx& - \alpha_{22}^2 \left| \alpha_{31}^{} \right| \alpha_{33}^{} s_{12}^{} c_{12}^{} s_{23}^{} c_{23}^2 \sin\phi_{31}^{} + \alpha_{22}^2 \left| \alpha_{32}^{} \right| \alpha_{33}^{} s_{12}^2 s_{23}^{} c_{23}^{} \sin\phi_{32}^{}\,,\\
		\mathcal{V}_{\mu\tau}^{23} &\approx& + \alpha_{22}^2 \left| \alpha_{31}^{} \right| \alpha_{33}^{} s_{12}^{} c_{12}^{} s_{23}^{} c_{23}^2 \sin\phi_{31}^{} + \alpha_{22}^2 \left| \alpha_{32}^{} \right| \alpha_{33}^{} c_{12}^2 s_{23}^{} c_{23}^{} \sin\phi_{32}^{}\,.\label{eq:Vmt230}
\end{eqnarray}
It is evident that the CP asymmetries ${\cal A}^{}_{\alpha \beta}$ in Eq.~(\ref{ACPnonuni}) are nonzero even with the trivial values of CP-violating phases $\{\rho, \sigma, \delta\}$. Except for ${\cal A}^{}_{ee}$, all those CP asymmetries are purely induced by the non-unitary parameters.
\end{itemize}

\subsection{Minimal Seesaw Model}
\subsubsection{CP Asymmetries}
In the previous discussions, one can recognize that there are quite a number of parameters in the calculations of CP asymmetries ${\cal A}^{}_{\alpha \beta}$. In order to simplify the situation, we consider the so-called minimal seesaw model, which extends the SM with two right-handed neutrino singlets~\cite{Kleppe:1995zz, Ma:1998zg, King:1999mb, King:2002nf, Frampton:2002qc, Guo:2006qa, Xing:2020ald}. One salient feature of the minimal seesaw model is that the lightest neutrino mass is vanishing, namely, $m_1^{} = 0$ in the NO case or $m_3^{} = 0$ in the IO case. Meanwhile, as the lightest neutrino is massless, there exists only one Majorana-type CP-violating phase. For definiteness, we keep the Majorana-type CP-violating phase $\sigma$ in either NO or IO case. It should be kept in mind that $\sigma$ refers to the relative phase between the neutrino mass eigenstates $|\nu^{}_2\rangle$ and $|\nu^{}_3\rangle$ in the NO case, whereas that between $|\nu^{}_2\rangle$ and $|\nu^{}_1\rangle$ in the IO case. In practice, we can simply set $\rho = 0$ in the standard parametrization of $\widetilde{V}$ in Eq.~(\ref{eq:standardpara}).

In the NO case with $m_1^{} = 0$, the expressions of CP-violating asymmetries ${\cal A}^{}_{\alpha \beta}$ in Eq.~(\ref{ACPnonuni}) are reduced to
\begin{eqnarray}
\mathcal{A}_{\alpha \beta}^{} = \frac{2 \mathcal{V}_{\alpha\beta}^{23} \sin F_{32}^{}}{\displaystyle \left| \sqrt{ \frac{m_2^{}}{m_3^{}} } V_{\alpha 2}^{} V_{\beta 2}^{} + \sqrt{ \frac{m_3^{}}{m_2^{}} } V_{\alpha 3}^{} V_{\beta 3}^{} \right|^2_{}- 4\, \mathcal{C}_{\alpha\beta}^{23} \sin^2_{} \frac{F_{32}^{}}{2}} \, .
\label{eq:AabNO}
\end{eqnarray}	
Given $m_1^{} = 0$, one can figure out $m_2^{}/m^{}_3 = \sqrt{\Delta m_{21}^2/\Delta m^2_{31}} \approx 0.172$, where the best-fit values $\Delta m_{21}^2 = 7.42 \times 10^{-5}_{}\, {\rm eV^2}$ and $\Delta m_{31}^2 = 2.51 \times 10^{-3}_{}\, {\rm eV^2}$ have been input~\cite{Esteban:2020cvm}. In the IO case with $m_3^{}=0$, the CP asymmetries are given by
\begin{eqnarray}
\mathcal{A}_{\alpha \beta}^{} &=& \frac{2 \mathcal{V}_{\alpha\beta}^{12} \sin F_{21}^{}}{\displaystyle \left| \sqrt{\frac{m_1^{}}{m^{}_2}} V_{\alpha 1}^{} V_{\beta 1}^{} + \sqrt{\frac{m_2^{}}{m^{}_1}} V_{\alpha 2}^{} V_{\beta 2}^{} \right|^2_{}- 4\, \mathcal{C}_{\alpha\beta}^{12}\sin^2_{} \frac{F_{21}^{}}{2}} \,,
\label{eq:AabIO}
\end{eqnarray}	
where the neutrino mass ratio can be estimated as $m^{}_1/m^{}_2 = \sqrt{(\Delta m^2_{32} + \Delta m^2_{21})/\Delta m^2_{32}} \approx 0.985$ for the best-fit values $\Delta m_{21}^2 = 7.42 \times 10^{-5}_{}\, {\rm eV^2}$ and $\Delta m_{32}^2 = -2.50 \times 10^{-3}_{}\, {\rm eV^2}$. Note that two nonzero neutrino masses $m^{}_1$ and $m^{}_2$ are nearly degenerate in the IO case. Once the other mixing parameters in $V^{}_{\alpha i}$ are known in either NO or IO case, one can easily calculate the CP asymmetries.

\subsubsection{Numerical Results}
We proceed with a numerical illustration for the CP asymmetries in the neutrino-antineutrino oscillations with a non-unitary mixing matrix $V$ in the framework of minimal seesaw model. First of all, we have to specify the input values of all the parameters for our numerical calculations. The CP asymmetries ${\cal A}^{}_{\alpha \beta}$ depend on the mixing matrix $V = T\cdot \widetilde{V}$, for which we adopt the triangular parametrization and impose the latest constraints on the unitarity violation~\cite{Fernandez-Martinez:2016lgt} characterized by the triangular matrix $T$. The $2\sigma$ allowed ranges of relevant non-unitary parameters have been given in Eqs.~(\ref{eq:alphadiag}) and (\ref{eq:alphaoff}), whereas three phases $\{\phi^{}_{21}, \phi^{}_{31}, \phi^{}_{32}\}$ are completely free. In addition, the best-fit values for neutrino oscillation parameters from Ref.~\cite{Esteban:2020cvm}, i.e., Eqs.~(\ref{eq:mixparaNO}) and (\ref{eq:mixparaIO}), will be used in the NO and IO case, respectively.
To clearly show the impact of non-unitary parameters on the CP asymmetries, we define the working observable
\begin{eqnarray}
\varepsilon_{\alpha\beta}^{} \equiv \frac{\mathcal{A}_{\alpha \beta}^{} - \widetilde{\mathcal{A}}_{\alpha \beta}^{}}{\widetilde{\mathcal{A}}_{\alpha \beta}^{}} \times 100\% \,,
\label{eq:varepsi}
\end{eqnarray}
where the CP asymmetries in the unitary limit are denoted by $\widetilde{\mathcal{A}}_{\alpha \beta}^{}$. It is worthwhile to mention that $\widetilde{\cal A}^{}_{\alpha \beta}$ are computed in the same way as ${\cal A}^{}_{\alpha \beta}$ in Eq.~(\ref{eq:AabNO}) or Eq.~(\ref{eq:AabIO}) but with $T = {\bf 1}$ or equivalently $V = \widetilde{V}$. In the NO case, we further fix the oscillation phase at $F^{}_{32} = \pi/2$ and vary the non-unitary parameters. When ${\cal A}^{}_{\alpha \beta}$ reach their maxima ${\cal A}^{\rm max}_{\alpha \beta}$ or minima ${\cal A}^{\rm min}_{\alpha \beta}$, we accordingly obtain the upper limits $\varepsilon^{\rm U}_{\alpha \beta} \equiv ({\cal A}^{\rm max}_{\alpha \beta} - \widetilde{\cal A}^{}_{\alpha \beta})/\widetilde{\cal A}^{}_{\alpha \beta}$ or the lower limits $\varepsilon^{\rm L}_{\alpha \beta} \equiv ({\cal A}^{\rm min}_{\alpha \beta} - \widetilde{\cal A}^{}_{\alpha \beta})/\widetilde{\cal A}^{}_{\alpha \beta}$. In a similar way, $F^{}_{21} = \pi/2$ is taken in the IO case, and $\varepsilon^{\rm U}_{\alpha \beta}$ and $\varepsilon^{\rm L}_{\alpha \beta}$ can be found. The final results in the NO and IO cases are summarized in Table~\ref{tab:str} for $\sigma = 0^\circ_{}$ and $\sigma = 45^\circ_{}$, where some interesting observations can be made.

\begin{table}
\centering
\begin{tabular}{c|p{0.12\textwidth}|p{0.12\textwidth}|p{0.12\textwidth}|c}
\toprule \hline
\multirow{2}{*}{Normal Ordering} & \multicolumn{2}{c|}{$\delta = 195^\circ_{}$, $\sigma=0^\circ_{}$} & \multicolumn{2}{c}{$\delta = 195^\circ_{}$, $\sigma=45^\circ_{}$} \\
\cline{2-3}
\cline{4-5}
& $\makecell[c]{\varepsilon^{\rm U}_{\alpha\beta}}$ & $\makecell[c]{\varepsilon^{\rm L}_{\alpha\beta}}$ & $\makecell[c]{\varepsilon^{\rm U}_{\alpha\beta}}$ & $\makecell[c]{\varepsilon^{\rm L}_{\alpha\beta}}$ \\
		\hline
		$\alpha$, $\beta$ = $e$, $e$ & $\makecell[c]{0\%}$ & $\makecell[c]{0\%}$ & $\makecell[c]{0\%}$ & $\makecell[c]{0\%}$ \\
		\hline
		$\alpha$, $\beta$ = $e$, $\mu$ & $\makecell[c]{-0.008974\%}$ & $\makecell[c]{+0.008974\%}$ & $\makecell[c]{-0.001717\%}$ & $\makecell[c]{+0.001717\%}$ \\
		\hline
		$\alpha$, $\beta$ = $e$, $\tau$ & $\makecell[c]{+1.946\%}$ & $\makecell[c]{-1.948\%}$ & $\makecell[c]{+0.2681\%}$ & $\makecell[c]{-0.2698\%}$ \\
		\hline
		$\alpha$, $\beta$ = $\mu$, $\mu$ & $\makecell[c]{+0.09932\%}$ & $\makecell[c]{-0.09932\%}$ & $\makecell[c]{+0.005116\%}$ & $\makecell[c]{-0.005116\%}$ \\
		\hline
		$\alpha$, $\beta$ = $\mu$, $\tau$ & $\makecell[c]{-206.8\%}$ & $\makecell[c]{+206.8\%}$ & $\makecell[c]{-0.4555\%}$ & $\makecell[c]{+0.4564\%}$ \\
		\hline
		$\alpha$, $\beta$ = $\tau$, $\tau$ & $\makecell[c]{-19.39\%}$ & $\makecell[c]{+19.40\%}$ & $\makecell[c]{+0.9050\%}$ & $\makecell[c]{-0.9000\%}$ \\
		\hline \hline
		\multirow{2}{*}{Inverted Ordering} &
\multicolumn{2}{c|}{$\delta = 286^\circ_{}$, $\sigma=0^\circ_{}$} &
\multicolumn{2}{c}{$\delta = 286^\circ_{}$, $\sigma=45^\circ_{}$}\\
\cline{2-3}
\cline{4-5}
& $\makecell[c]{\varepsilon^{\rm U}_{\alpha\beta}}$ & $\makecell[c]{\varepsilon^{\rm L}_{\alpha\beta}}$ & $\makecell[c]{\varepsilon^{\rm U}_{\alpha\beta}}$ & $\makecell[c]{\varepsilon^{\rm L}_{\alpha\beta}}$ \\
		\hline
		$\alpha$, $\beta$ = $e$, $e$ & $\makecell[c]{0\%}$ & $\makecell[c]{0\%}$ & $\makecell[c]{0\%}$ & $\makecell[c]{0\%}$ \\
		\hline
		$\alpha$, $\beta$ = $e$, $\mu$ & $\makecell[c]{+0.02049\%}$ & $\makecell[c]{-0.02049\%}$ & $\makecell[c]{+0.002801\%}$ & $\makecell[c]{-0.002801\%}$ \\
		\hline
		$\alpha$, $\beta$ = $e$, $\tau$ & $\makecell[c]{-2.998\%}$ & $\makecell[c]{+2.994\%}$ & $\makecell[c]{+0.2408\%}$ & $\makecell[c]{-0.2483\%}$ \\
		\hline
		$\alpha$, $\beta$ = $\mu$, $\mu$ & $\makecell[c]{-0.01942\%}$ & $\makecell[c]{+0.01942\%}$ & $\makecell[c]{-0.01444\%}$ & $\makecell[c]{+0.01444\%}$ \\
		\hline
		$\alpha$, $\beta$ = $\mu$, $\tau$ & $\makecell[c]{-12.13\%}$ & $\makecell[c]{+12.02\%}$ & $\makecell[c]{-0.5561\%}$ & $\makecell[c]{+0.5401\%}$ \\
		\hline
		$\alpha$, $\beta$ = $\tau$, $\tau$ & $\makecell[c]{+3.029\%}$ & $\makecell[c]{-3.028\%}$ & $\makecell[c]{-1.597\%}$ & $\makecell[c]{+1.626\%}$ \\
		\hline
		\bottomrule
\end{tabular}
\vspace{0.3cm}
\caption{The deviations of the CP asymmetries ${\cal A}^{}_{\alpha \beta}$ with the non-unitary mixing matrix from those $\widetilde{\cal A}^{}_{\alpha \beta}$ in the unitary limit, where $\varepsilon^{\rm U}_{\alpha \beta} \equiv ({\cal A}^{\rm max}_{\alpha \beta} - \widetilde{\cal A}^{}_{\alpha \beta})/\widetilde{\cal A}^{}_{\alpha \beta}$ and $\varepsilon^{\rm L}_{\alpha \beta} \equiv ({\cal A}^{\rm min}_{\alpha \beta} - \widetilde{\cal A}^{}_{\alpha \beta})/\widetilde{\cal A}^{}_{\alpha \beta}$ have been obtained by varying the non-unitary parameters within their $2\sigma$ ranges in Eqs.~(\ref{eq:alphadiag}) and (\ref{eq:alphaoff}), as well as the free phases $\{\phi^{}_{21}, \phi^{}_{31}, \phi^{}_{32}\}$. In the NO case, the oscillation phase $F^{}_{32} = \pi/2$ is taken and the best-fit values for neutrino mixing parameters in Eq.~(\ref{eq:mixparaNO}) are input. In the IO case, we fix the oscillation phase $F^{}_{21} = \pi/2$ and take the best-fit values for neutrino mixing parameters in Eq.~(\ref{eq:mixparaIO}). In addition, $\sigma = 0^\circ$ and $\sigma = 45^\circ$ have been chosen for illustration.}
\label{tab:str}
\end{table}

\begin{itemize}
\item From Table~\ref{tab:str}, one can observe that $\varepsilon^{\rm U}_{ee} = \varepsilon^{\rm L}_{ee} = 0$, namely, the CP asymmetry ${\cal A}^{}_{ee}$ is not affected by unitarity violation. In order to understand this feature, we take $\alpha = \beta = e$ in Eq.~(\ref{ACPnonuni}) and then arrive at
\begin{eqnarray}
\mathcal{A}_{ee}^{} = \frac{\displaystyle \sum_{i<j} {\rm Im}\left[V_{ei}^2 V_{ej}^{\ast 2}\right] m_i^{} m_j^{}\sin F_{ji}^{} }{\displaystyle \left|\sum_{i=1}^3 V_{ei}^2 m_i^{} \right|^2 - 4 \sum_{i<j} {\rm Re}\left[V_{ei}^2 V_{ej}^{\ast 2}\right] m_i^{} m_j^{} \sin^2 \frac{F_{ji}^{}}{2}} \, ,	
\label{Aee}
\end{eqnarray}
where the non-unitary mixing matrix is given by $V = T\cdot \widetilde{V}$. More explicitly, we have $V_{ei}^{} = \alpha_{11}^{} \widetilde{V}_{ei}^{}$, indicating that the overall factor $\alpha^4_{11}$ in the numerator and the denominator of Eq.~(\ref{Aee}) will be exactly cancelled out. Consequently, one obtains $\mathcal{A}_{ee}^{} = \widetilde{\mathcal{A}}_{ee}^{}$, where $\widetilde{\cal A}^{}_{ee}$ is calculated by using the unitary mixing matrix $\widetilde{V}$.

\item Comparing the expressions of $\mathcal{V}_{\alpha\beta}^{ij}$ with those of $\widetilde{\mathcal{V}}_{\alpha\beta}^{ij}$, one finds that the terms $\alpha^2_{\alpha \alpha} \alpha^2_{\beta \beta} \widetilde{\cal V}^{ij}_{\alpha \beta}$, where the subscripts of $\alpha^{}_{\beta\beta}$ for $\beta = e, \mu, \tau$ should be identified as $(e, \mu, \tau) = (1, 2, 3)$, play the leading role in $\mathcal{V}_{\alpha\beta}^{ij}$. Since $\alpha_{ii}^{} < 1$ holds for $i=1,2,3$, the absolute value of $\alpha_{\alpha\alpha}^2 \alpha_{\beta\beta}^2 \widetilde{\cal V}_{\alpha\beta}^{ij}$ will always be smaller than that of $\widetilde{\cal V}_{\alpha\beta}^{ij}$. If only such leading-order terms were taken into account, $\varepsilon^{}_{\alpha \beta}$ would be negative. However, this is not the case, since the contributions from the terms associated with $\left|\alpha_{ij}^{}\right|$ and $\phi_{ij}^{}$ (for $ij = 21, 31, 32$) may also be important. For this reason, the sign of $\varepsilon^{}_{\alpha\beta}$ can be either positive or negative due to the interplay between different contributions.

\item It should be noticed that $|\varepsilon^{\rm U}_{\mu\tau}| = |\varepsilon^{\rm L}_{\mu\tau}| = 206.8\%$ in the NO case with $\sigma = 0^\circ$ is remarkably larger than others. Such an observation can be understood by examining the approximate formula of $\varepsilon^{}_{\mu \tau}$, i.e.,
\begin{eqnarray}
\varepsilon_{\mu\tau}^{} &\approx& (\alpha^2_{22} \alpha^2_{33} - 1) - \alpha^2_{22} \alpha^{}_{33} c_{12}^{} s_{23}^{} c_{23}^{} \frac{\left|\alpha_{31}^{}\right| s_{12}^{} c_{23}^{} \sin \phi_{31}^{} + \left| \alpha_{32}^{} \right| c_{12}^{}  \sin \phi_{32}^{}}{{\cal J} \cos 2\theta^{}_{23}} \, , \quad \quad
\label{eq:epmut}
\end{eqnarray}	
where Eq.~(\ref{eq:emt23}) with $\sigma=0^\circ_{}$ and $\widetilde{\cal V}^{23}_{\mu\tau}$ from Appendix~\ref{app:UN} have been used. With the best-fit values $\delta = 195^\circ$ (i.e., $\sin \delta \approx -0.26$) and $\theta^{}_{23} = 49^\circ$ (i.e., $\cos 2\theta^{}_{23} \approx -0.14$), one can see that the denominator of the second term on the right-hand side of Eq.~(\ref{eq:epmut}) could be comparable to or even smaller than the numerator. However, this observation is not applicable to $\varepsilon^{}_{\alpha \beta}$ in other flavors.

\item One may also notice that $\varepsilon_{\alpha \beta}^{\rm U}$ and $\varepsilon^{\rm L}_{\alpha \beta}$ in the $e\mu$ and $\mu\mu$ flavors are much smaller than those in other flavors in both NO and IO cases, no matter whether $\sigma = 0^\circ_{}$ or $45^\circ_{}$ is assumed. From the analytical formulas of $\epsilon^{ij}_{\alpha \beta}$ in Eqs.~(\ref{eq:emm12})-(\ref{eq:emt23}), we can see that $\epsilon^{ij}_{e\mu}$ and $\epsilon^{ij}_{\mu \mu}$ are determined partially by the factors $(1 - \alpha_{11}^2 \alpha_{22}^2)$ and $(1 - \alpha_{22}^4)$, which are smaller than $(1 - \alpha_{11}^2 \alpha_{33}^2)$, $(1 - \alpha_{22}^2 \alpha_{33}^2)$ and $(1 - \alpha_{33}^4)$, since $\alpha_{22}^{}$ and $\alpha_{11}^{}$ are much closer to one than $\alpha_{33}^{}$ is. As for the contributions from the off-diagonal elements of $T$, the remaining terms in $\epsilon_{e\mu}^{ij}$ and $\epsilon_{\mu\mu}^{ij}$ are proportional to $|\alpha_{21}^{}|$, which is smaller by two orders of magnitude than $|\alpha_{31}^{}|$ and $|\alpha_{32}^{}|$. The latter two parameters appear in $\epsilon_{e\tau}^{ij}$, $\epsilon_{\mu\tau}^{ij}$ and $\epsilon_{\tau\tau}^{ij}$.
\end{itemize}

As we have briefly mentioned before, even if the ordinary CP-violating phases $\{\delta, \rho, \sigma\}$ take trivial values, there still exists CP violation in neutrino-antineutrino oscillations due to the non-unitary CP phases. In this special case with $\delta = \sigma = 0^\circ$, the CP asymmetries $\widetilde{\cal A}^{}_{\alpha \beta}$ in the minimal seesaw model are accordingly vanishing, so the definitions of $\varepsilon^{}_{\alpha \beta}$ in Eq.~(\ref{eq:varepsi}) are no longer valid and we just compute the maxima and minima of ${\cal A}^{}_{\alpha \beta}$. The numerical results are presented in Table~\ref{tab:000}, where the input values are the same as in Table~\ref{tab:str}, except for $\delta$ and $\sigma$. Some comments on the results in Table~\ref{tab:000} are in order. In both NO and IO cases, one can observe that the CP asymmetries ${\cal A}^{}_{e\mu}$ and ${\cal A}^{}_{\mu\mu}$ are much smaller than others, which are all at the $10^{-3}$ level. The main reason for such a significant suppression is the fact that the experimental upper limit on $|\alpha^{}_{21}|$ (i.e., $\lesssim 10^{-5}$) is smaller by two orders of magnitude than that on $|\alpha^{}_{31}|$ or $|\alpha^{}_{32}|$ (e.g., $\lesssim 10^{-3}$). For example, we consider the CP asymmetry ${\cal A}^{}_{e\mu}$ in the NO case, which is mainly determined by the Jarlskog-like parameter $\mathcal{V}_{e\mu}^{23}$. As indicated in Eq.~(\ref{eq:Vee230}), it is actually vanishing at the leading order. The first-order correction gives rise to $\mathcal{V}^{23}_{e\mu} \approx \alpha_{11}^2 \left|\alpha_{21}^{}\right| \alpha_{22}^{} s_{12}^2 s_{13}^{} c_{13}^3 s_{23}^{} \sin \phi_{21}^{}$, which is doubly suppressed by $|\alpha^{}_{21}|$ and $s^{}_{13}$, leading to ${\cal A}^{}_{e\mu} \sim 10^{-5}$ of the right order as shown in Table~\ref{tab:000}. In a similar way, one can understand the suppression of ${\cal A}^{}_{e\mu}$ in the IO case, and ${\cal A}^{}_{\mu\mu}$ in both NO and IO cases.
\begin{table}
\centering
\begin{tabular}{c|p{0.15\textwidth}|p{0.15\textwidth}|p{0.15\textwidth}|c}
\toprule \hline
\multirow{2}{*}{CP Asymmetries} &
\multicolumn{2}{c|}{Normal Ordering} & \multicolumn{2}{c}{Inverted Ordering}\\
\cline{2-3}
\cline{4-5}
& $\makecell[c]{\mathcal{A}^{\rm max}_{\alpha\beta}}$ & $\makecell[c]{\mathcal{A}^{\rm min}_{\alpha\beta}}$ & $\makecell[c]{\mathcal{A}^{\rm max}_{\alpha\beta}}$ & $\makecell[c]{\mathcal{A}^{\rm min}_{\alpha\beta}}$ \\
		\hline
		$\alpha$, $\beta$ = $e$, $e$ & $\makecell[c]{0}$ & $\makecell[c]{0}$ & $\makecell[c]{0}$ & $\makecell[c]{0}$ \\
		\hline
		$\alpha$, $\beta$ = $e$, $\mu$ & $\makecell[c]{+1.554\times 10^{-5}_{}}$ & $\makecell[c]{-1.554\times 10^{-5}_{}}$ & $\makecell[c]{+6.652\times 10^{-5}_{}}$ & $\makecell[c]{-6.652\times 10^{-5}_{}}$ \\
		\hline
		$\alpha$, $\beta$ = $e$, $\tau$ & $\makecell[c]{+4.676\times 10^{-3}_{}}$ & $\makecell[c]{-4.676\times 10^{-3}_{}}$ & $\makecell[c]{+9.177\times 10^{-3}_{}}$ & $\makecell[c]{-9.177\times 10^{-3}_{}}$ \\
		\hline
		$\alpha$, $\beta$ = $\mu$, $\mu$ & $\makecell[c]{+6.422\times 10^{-6}_{}}$ & $\makecell[c]{-6.422\times 10^{-6}_{}}$ & $\makecell[c]{+1.406\times 10^{-4}_{}}$ & $\makecell[c]{-1.406\times 10^{-4}_{}}$ \\
		\hline
		$\alpha$, $\beta$ = $\mu$, $\tau$ & $\makecell[c]{+1.187\times 10^{-3}_{}}$ & $\makecell[c]{-1.187\times 10^{-3}_{}}$ & $\makecell[c]{+6.756\times 10^{-3}_{}}$ & $\makecell[c]{-7.358\times 10^{-3}_{}}$ \\
		\hline
		$\alpha$, $\beta$ = $\tau$, $\tau$ & $\makecell[c]{+3.744\times 10^{-3}_{}}$ & $\makecell[c]{-3.744\times 10^{-3}_{}}$ & $\makecell[c]{+8.713\times 10^{-3}_{}}$ & $\makecell[c]{-8.713\times 10^{-3}_{}}$ \\
\bottomrule
\end{tabular}
\vspace{0.3cm}
\caption{The maxima and minima of the CP asymmetries ${\cal A}^{}_{\alpha \beta}$ in neutrino-antineutrino oscillations in the framework of minimal seesaw model with $\delta = \sigma = 0^\circ$, where the other input values are the same as in Table~\ref{tab:str}. }
\label{tab:000}
\end{table}

\begin{figure}[t!]
	\centering
	\includegraphics[scale=0.67]{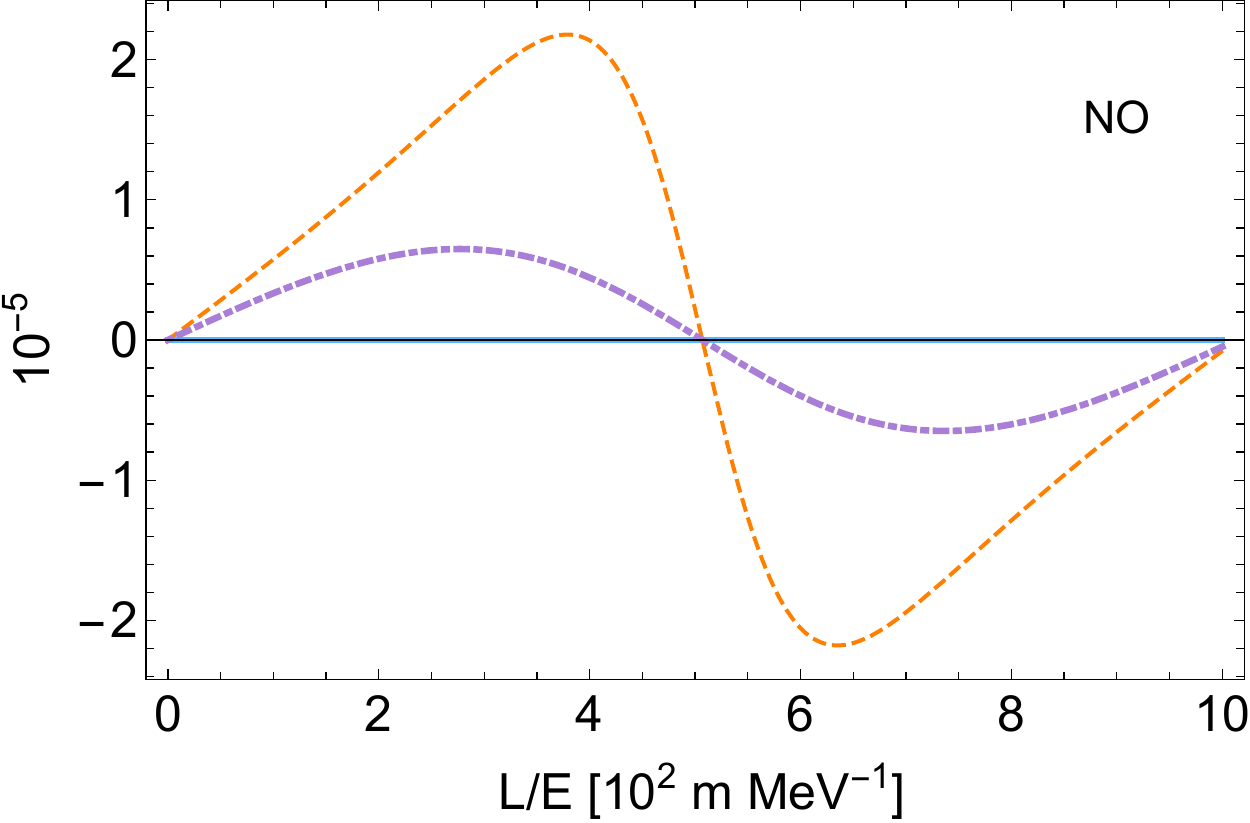}\quad
	\includegraphics[scale=0.67]{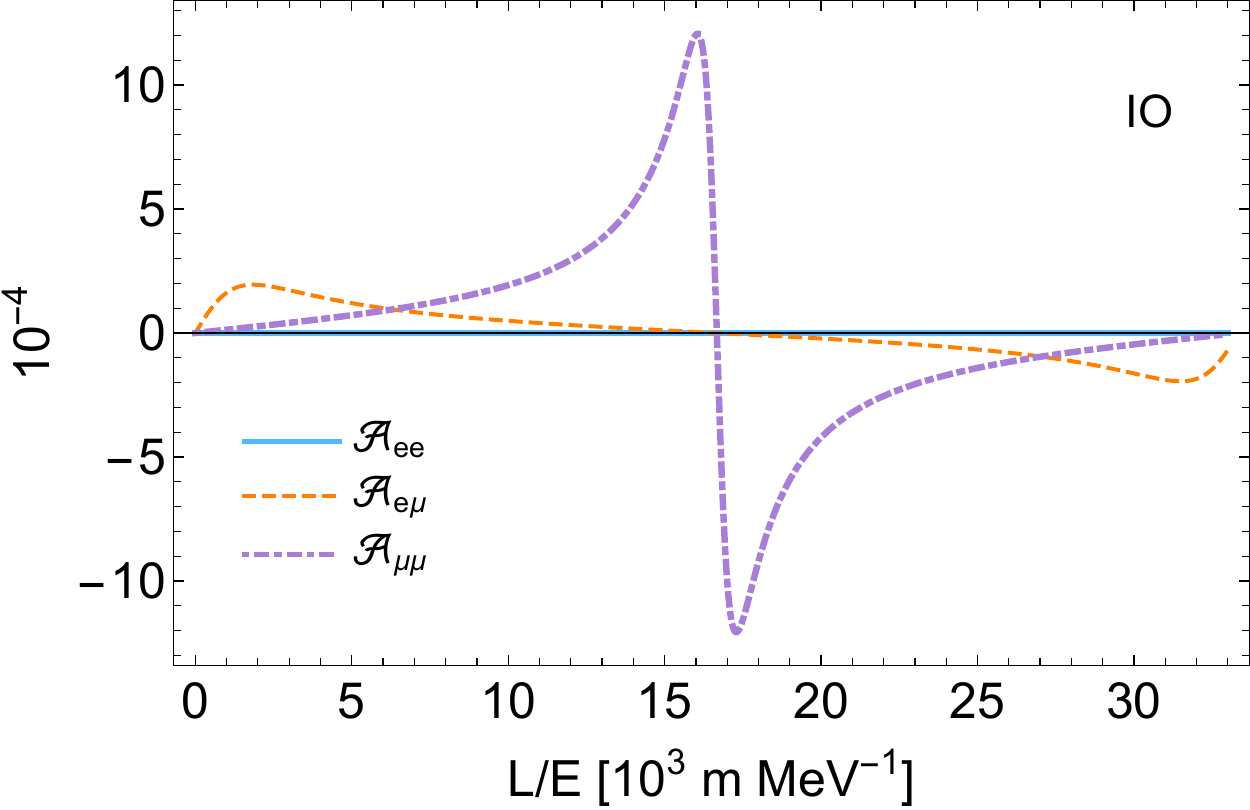}\\
\vspace{0.4cm}
	\includegraphics[scale=0.67]{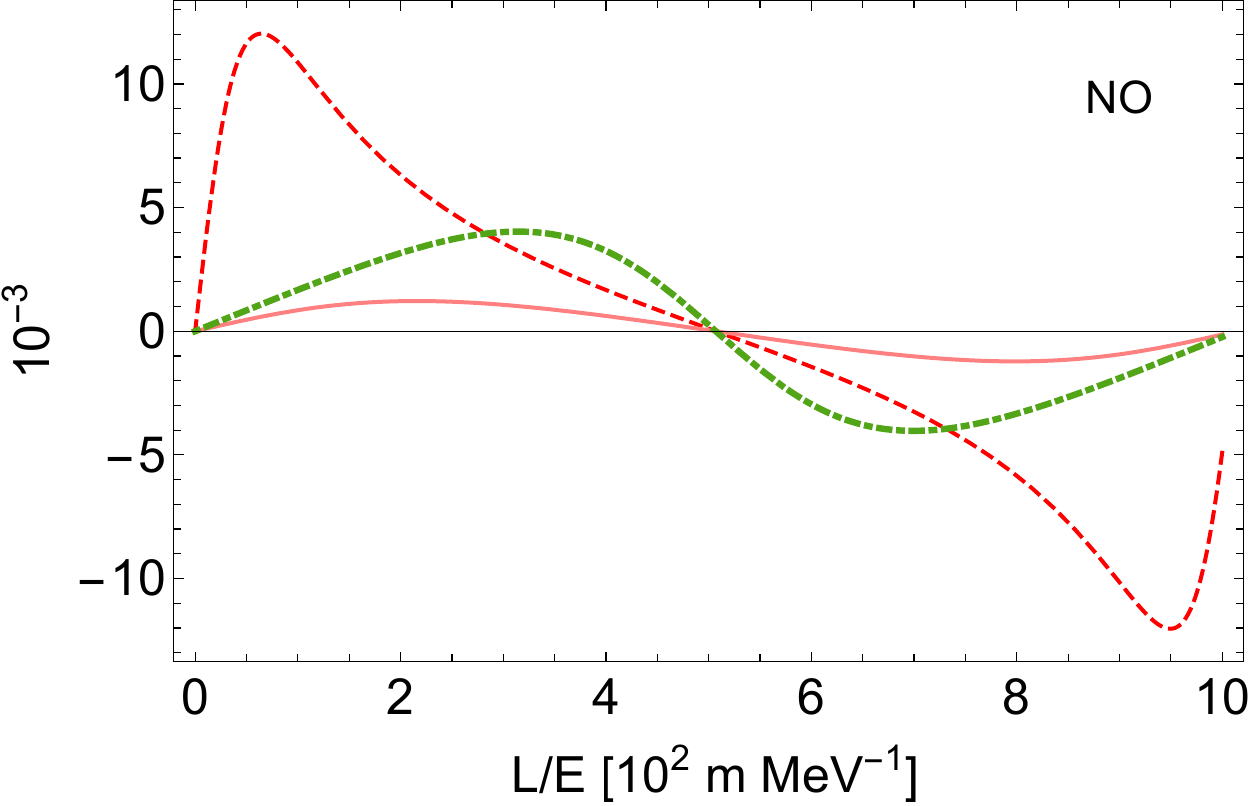}\quad
	\includegraphics[scale=0.67]{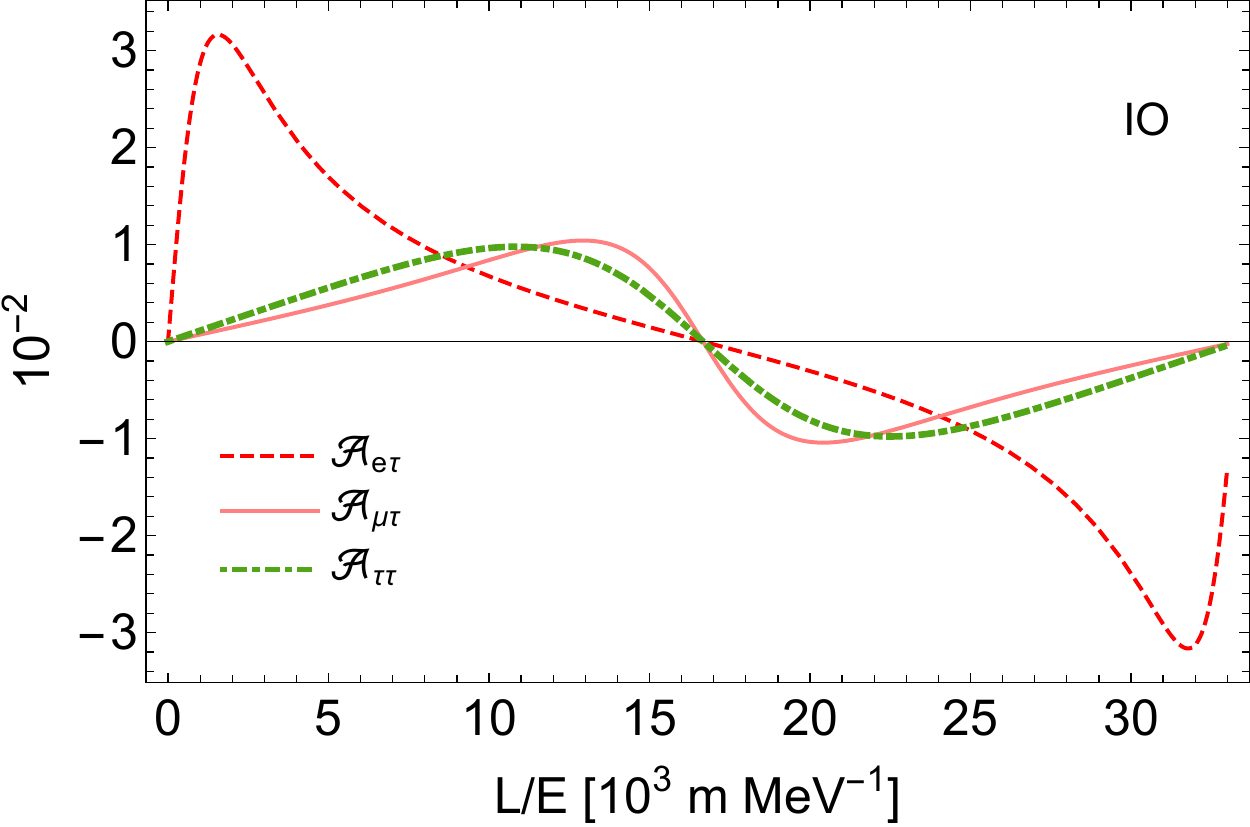}
\vspace{-0.3cm}
	\caption{Illustration for the dependence of CP asymmetries ${\cal A}^{}_{\alpha \beta}$ on $L/E$, where $L$ is the baseline length and $E$ is the neutrino beam energy. Note that the mixing parameters are the same as those adopted in Table~\ref{tab:000}, and three non-unitary CP phases $\{\phi_{21}^{}, \phi^{}_{31}, \phi^{}_{32}\}$ take the values at which ${\cal A}^{}_{\alpha \beta} = {\cal A}^{\rm max}_{\alpha \beta}$ in Table~\ref{tab:000} are reached.}
	\label{fig:fig1}
\end{figure}

In the foregoing numerical calculations, we have fixed the oscillation phase $F^{}_{32} = \pi/2$ and $F^{}_{21} = \pi/2$ in the NO and IO case, respectively. It will be interesting to see how the CP asymmetries depend on $L/E$, where $L$ is the baseline length and $E$ is the neutrino beam energy. The final results are shown in Fig.~ \ref{fig:fig1}, where two plots in the left column refer to the NO case while those in the right column to the IO case. Except for the oscillation phase, the other input values for relevant parameters remain the same as in Table~\ref{tab:000}. In addition, the non-unitary phases $\{\phi^{}_{21}, \phi^{}_{31}, \phi^{}_{32}\}$ take the values when the maxima ${\cal A}^{\rm max}_{\alpha \beta}$ in Table~\ref{tab:000} are reached. Most curves for the CP asymmetries ${\cal A}^{}_{\alpha \beta}$ in Fig.~\ref{fig:fig1} show sizable deviations from the sinusoidal shape, as the denominator of Eq.~(\ref{eq:AabNO}) or Eq.~(\ref{eq:AabIO}) depends on $\sin^2 (F^{}_{32}/2)$ or $\sin^2 (F^{}_{21}/2)$. Therefore, given the non-unitary phases and other mixing parameters, one can choose an optimal value of $L/E$ to maximize the CP asymmetry ${\cal A}^{}_{\alpha \beta}$ for a specific oscillation channel.

\subsection{Heavy Majorana Neutrinos}\label{subsec:hna}

Thus far we have focused only on the neutrino-antineutrino oscillations of three light Majorana neutrinos, for which the oscillation probabilities are highly suppressed by the small ratios $m^2_i/E^2$ and the sizes of CP asymmetries associated with the non-unitary mixing matrix are limited by current experimental constraints on non-unitary mixing parameters. In this subsection, we make a brief comparison between the neutrino-antineutrino oscillations for light Majorana neutrinos and those for heavy Majorana neutrinos in the minimal seesaw model.

In Ref.~\cite{Antusch:2020pnn}, the collider signals of heavy Majorana neutrinos $N^{}_i$ in the seesaw model have been examined and the neutrino-antineutrino oscillations of heavy Majorana neutrinos have been studied in the framework of quantum field theories. The main idea is to probe the displaced vertices induced by heavy Majorana neutrinos that are produced in the lepton-number-violating (LNV) processes $W^+ \to l^+_\alpha N^{}_i \to l^+_\alpha l^+_\beta j j$ and in the lepton-number-conserving (LNC) processes $W^+ \to l^+_\alpha N^{}_i \to l^+_\alpha l^-_\beta j j$, where the initial virtual gauge boson $W^+$ is produced in the large hadron colliders. The heavy neutrino-antineutrino oscillations could lead to an oscillating rate of LNV and LNC events with respect to the distance between the production and decay vertices of heavy Majorana neutrinos~\cite{Antusch:2020pnn}. In the minimal seesaw model, for two heavy Majorana neutrinos of an averaged mass $(M^{}_1 + M^{}_2)/2 = 7~{\rm GeV}$ and a tiny mass splitting determined by $M^2_2 - M^2_1 = 1.04\times 10^{-11}~{\rm GeV}^2$, the neutrino-antineutrino oscillation length is estimated to be $L^{}_{\rm osc} \approx 8.34~{\rm cm}$ for the Lorentz factor $\gamma \equiv E/M^{}_i \approx 50$. Furthermore, given that the uncertainties in the measurements of the momenta of the final-state charged leptons and jets are respectively $(0.5\% - 1\%)$ and $(5\% - 30\%)$ at the detectors, the neutrino-antineutrino oscillations can indeed be coherent. It has been found in Ref.~\cite{Antusch:2020pnn} that the probability of heavy neutrino-antineutrino oscillations in the $\mu \mu$-channel can reach $0.4$ for the relevant CP-violating Majorana phase being fixed at $\pi/3$. Notice that this probability has been obtained under the assumption that the spin correlation between the particles at production and decay vertices can be neglected~\cite{Antusch:2020pnn}, so there is no suppression by $M^2_i/E^2 = \gamma^{-2}$ at all. If the spin correlation is taken into account, the suppression factor $\gamma^{-2} \approx 4\times 10^{-4}$ is expected. However, compared to $m^2_i/E^2 \sim 10^{-14}$ in the case of light Majorana neutrinos, such a suppression is much less significant.

Regarding the CP asymmetries in the neutrino-antineutrino oscillations of three light Majorana neutrinos, they are also highly suppressed. In this connection, it is interesting to mention that the CP asymmetries induced by heavy Majorana neutrinos could be resonantly enhanced, as observed in Ref.~\cite{Bray:2007ru}. For two nearly-degenerate heavy Majorana neutrinos, i.e., $M_2^{} - M_1^{} \equiv \Delta M \sim \Gamma_i^{}$, where $\Gamma_i^{}$ (for $i = 1, 2$) represent the decay widths of two heavy Majorana neutrinos, the CP asymmetries between the LNV processes $q \bar{q}^\prime \to W^{+*} \to l^+_\alpha N^{}_i \to l^+_\alpha l^+_\beta W^-$ and their CP-conjugate processes $\bar{q} q^\prime \to W^{-*} \to l^-_\alpha N^{}_i \to l^-_\alpha l^-_\beta W^+$ at the CERN large hadron collider (LHC) could be large. As demonstrated in Ref.~\cite{Bray:2007ru}, one of the most promising ways for observing the CP asymmetries induced by heavy Majorana neutrinos at the LHC is to detect $pp \to e^+ \mu^+ W^- X$ and its CP-conjugate process $pp \to e^- \mu^- W^+ X$, where $X$ denotes all possible final states and the gauge bosons $W^\pm$ are assumed to decay hadronically. The corresponding CP asymmetry can be defined as
\begin{eqnarray}
	{\cal A}_{\rm CP}^{N} = \frac{\sigma \left( pp \to e^+_{} \mu^+_{} W^- X \right) - \mathcal{K} \sigma \left( p p \to e^-_{} \mu^-_{} W^+ X \right)}{\sigma \left( pp \to e^+_{} \mu^+_{} W^- X \right) + \mathcal{K} \sigma \left( pp \to e^-_{} \mu^-_{} W^+ X \right)} \, ,
	\label{eq:CPN}
\end{eqnarray}
where $\mathcal{K}$ accounts for the difference in the production of $W^+_{}$ and $W^-_{}$ in the high-energy proton-proton collisions. For the mixing matrix $R$ involved in the charged-current interaction of heavy Majorana neutrinos in Eq.~(\ref{eq:unitary}), one can take the values of $R^{}_{\alpha i} = 0.05$ (for $\alpha = e, \mu, \tau$ and $i = 1, 2$), except for $R^{}_{\mu 2} = 0.05{\rm i}$ and $R^{}_{\tau 2} = -0.05$, as in {\bf Scenario 3} in Ref.~\cite{Bray:2007ru}. After imposing the kinematic cuts on the transverse momenta $p^{}_{\rm T} > 15~{\rm GeV}$ and on the pseudo-rapidity $|\eta| > 2.5$ globally for all final-state particles, the cross sections $\sigma \left( pp \to e^+_{} \mu^+_{} W^- X \right)$ and $\sigma \left( p p \to e^-_{} \mu^-_{} W^+ X \right)$ are found to be in the range of $[0.01, 10]~{\rm fb}$ for the heavy Majorana neutrino mass $M^{}_1 \in [100, 500]~{\rm GeV}$. Given $M_1^{} = 200~{\rm GeV}$ and $\Delta M \sim 0.3~{\rm GeV}$, the CP asymmetry defined in Eq.~(\ref{eq:CPN}) is ${\cal A}^N_{\rm CP} \sim -0.2$~\cite{Bray:2007ru}.

Although the non-unitarity induced CP asymmetries in neutrino-antineutrino oscillations of three light Majorana neutrinos are practically tiny, the counterparts in the sector of heavy Majorana neutrinos may be sizable. Furthermore, inspired by the neutrino-antineutrino oscillations of heavy Majorana neutrinos investigated in Ref.~\cite{Antusch:2020pnn} and the CP asymmetries in the LNV processes mediated by heavy Majorana neutrinos studied in Ref.~\cite{Bray:2007ru}, one may explore the CP asymmetries solely from heavy neutrino-antineutrino oscillations in the collider searches for heavy Majorana neutrinos. As the mixing matrices $V$ and $R$ are intimately correlated in the seesaw model, it is intriguing to carry out a global analysis of CP asymmetries in neutrino-antineutrino oscillations for both light and heavy Majorana neutrinos. Such an analysis will be left for a future work.

\section{Summary}\label{sec:sum}

In this paper, we have examined the CP asymmetries in the neutrino-antineutrino oscillations in the presence of a non-unitary flavor mixing matrix. The main motivation for such a study is two-fold. First, neutrino-antineutrino oscillations occur only when massive neutrinos are Majorana particles. As in a class of seesaw models, three light neutrinos turn out to be Majorana particles, and the flavor mixing matrix of three light neutrinos is intrinsically non-unitary. Second, the non-unitarity of the mixing matrix brings in extra sources of CP violation, which is quite different from the unitary case.

By using the QR factorization, we establish the relation between the Hermitian parametrization and the triangular parametrization of a non-unitary mixing matrix. Then, the CP asymmetries in neutrino-antineutrino oscillations with a non-unitary mixing matrix are found to be essentially governed by 18 independent Jarlskog-like parameters, whose analytical expressions at the leading order are derived. Finally, implementing the latest experimental constraints on the leptonic unitarity violation, we numerically compute the CP asymmetries in the minimal seesaw model, where the number of model parameters can be further reduced. It is worthwhile to stress that even with trivial values of ordinary CP-violating phases $\{\rho, \sigma, \delta\}$, one can obtain nonzero CP asymmetries due to the extra non-unitary CP phases $\{\phi^{}_{21}, \phi^{}_{31}, \phi^{}_{32}\}$.

Current interest in neutrino-antineutrino oscillations is basically academic, as the oscillation probabilities are highly suppressed by the squared mass-to-energy ratio $m^2_i/E^2$ (e.g., $\sim 10^{-14}$ for $m^{}_i \sim 0.1~{\rm eV}$ and $E \sim 1~{\rm MeV}$). However, if three ordinary neutrinos are indeed massive Majorana particles, then one has to experimentally measure two Majorana-type CP-violating phases as well. In addition, the probabilities of heavy neutrino-antineutrino oscillations may not be suppressed, as the masses of heavy Majorana neutrinos could be comparable to their momenta when they are produced in the large hadron colliders~\cite{Antusch:2020pnn}. For the CP asymmetries in the LNV processes induced by heavy Majorana neutrinos at the large hadron colliders, they can be resonantly enhanced if the mass splitting of two heavy Majorana neutrinos is on the same order of their decay widths~\cite{Bray:2007ru}. As for the long-term plan, great efforts will be made in the experimental detection of neutrino-antineutrino oscillations or other lepton-number- and CP-violating processes. Only in this way can one completely determine the fundamental parameters associated with massive Majorana neutrinos. In any case, our results on the basic properties of CP violation induced by leptonic non-unitarity will be helpful.

\section*{Acknowledgements}

The authors are indebted to Prof. Zhi-zhong Xing for helpful discussions. This work was supported in part by the National Natural Science Foundation of China under grant No. 11775232 and No. 11835013, by the Key Research Program of the Chinese Academy of Sciences under grant No. XDPB15, and by the CAS Center for Excellence in Particle Physics.

\appendix

\section{QR Factorization}
\label{app:QR}

In this Appendix, we present some details of the QR factorization of an arbitrary $3 \times 3$ complex matrix into the product of a unitary matrix $Q$ and an upper-triangular matrix $R$ with non-negative diagonal elements~\cite{Matrix}. As mentioned in Sec.~\ref{sec:para}, by using the QR factorization, one can establish the relationship between the Hermitian and triangular parametrizations of the non-unitary mixing matrix $V$. For the Hermitian parametrization of $V$ in Eq.~(\ref{eq:hermitianpara}), we can further decompose the Hermitian matrix $({\bf 1} - \eta)$ into a lower-triangular matrix and a unitary matrix by following the strategy outlined below.

First, the explicit expression of the Hermitian matrix $({\bf 1} - \eta)$ is given by
\begin{eqnarray}
	{\bf 1} - \eta = \begin{pmatrix}
		1 - \eta_{ee}^{} && -\eta_{e \mu}^{} && -\eta_{e \tau}^{} \\
		-\eta_{e \mu}^\ast && 1 - \eta_{\mu \mu}^{} && -\eta_{\mu \tau}^{} \\
		-\eta_{e \tau}^\ast && -\eta_{\mu \tau}^\ast && 1 - \eta_{\tau \tau}^{}
	\end{pmatrix} \;,
\end{eqnarray}
of which three columns will be denoted by three vectors ${\bf x}^{}_1$, ${\bf x}^{}_2$ and ${\bf x}^{}_3$. The first step is to find out a unitary matrix $U^{}_1$ such that $U^{}_1 \cdot {\bf x}^{}_1 = a^{}_1 {\bf e}^{}_1$, where $a^{}_1 = \left[(1 - \eta_{ee}^{})^2_{} + | \eta_{e \mu}^{} |^2_{} + | \eta_{e \tau}^{} |^2_{}\right]^{1/2} > 0$ is the modulus of the vector ${\bf x}^{}_1$ and ${\bf e}^{}_1 \equiv (1, 0, 0)^{\rm T}$ is the basis vector. In general, one can explicitly construct the unitary matrix $U^{}_1$ that transforms one given vector ${\bf x}^{}_1$ to another target vector ${\bf y}^{}_1$, i.e., $U^{}_1\cdot {\bf x}^{}_1 = {\bf y}^{}_1$. First, define the phase $\psi \equiv \arg \left({\bf x}^\dagger_1 \cdot {\bf y}^{}_1 \right) \in [0, 2\pi)$ and construct the auxiliary vector ${\bm \omega}^{}_1 \equiv e^{{\rm i}\psi} {\bf x}^{}_1 - {\bf y}^{}_1$. Then, the so-called Householder matrix is $U^{}_{{\bm \omega}^{}_1} \equiv {\bf 1} - 2({\bm \omega}^\dagger_1\cdot {\bm \omega}^{}_1)^{-1} {\bm \omega}^{}_1\cdot {\bm \omega}^\dagger_1$, and the desired unitary matrix is determined by $U^{}_1 = e^{{\rm i}\psi} U^{}_{{\bm \omega}^{}_1}$. One can easily prove that $U^{}_1 \cdot {\bf x}^{}_1 = {\bf y}^{}_1$ as expected. In our case, we have ${\bf x}^{}_1 \equiv (1 - \eta_{ee}^{}, -\eta_{e \mu}^\ast, -\eta_{e \tau}^\ast)^{\rm T}_{}$ and ${\bf y}^{}_1 = a^{}_1 {\bf e}^{}_1$, so the phase $\psi = 0$ is trivial and the unitary matrix $U^{}_1$ is found to be
\begin{eqnarray}
	U_1^{} = \frac{1}{a_1^{}} \begin{pmatrix}
		1 - \eta_{ee}^{} && -\eta_{e \mu}^{} && -\eta_{e \tau}^{} \\
		-\eta_{e \mu}^\ast && \displaystyle - 1 + \eta_{ee}^{} + \frac{|\eta_{e \tau}^{}|^2}{ a_1^{} - 1 + \eta_{ee}^{}} && \displaystyle
		-\frac{\eta_{e \mu}^\ast \eta_{e \tau}^{}}{a_1^{} - 1 + \eta_{ee}^{}} \\
		-\eta_{e \tau}^\ast && \displaystyle -\frac{\eta_{e \mu}^{} \eta_{e \tau}^\ast}{ a_1^{} - 1 + \eta_{ee}^{}} && \displaystyle - 1 + \eta_{ee}^{} + \frac{|\eta_{e \mu}^{}|^2}{ a_1^{} - 1 + \eta_{ee}^{}}
	\end{pmatrix}\;,
\end{eqnarray}
which is exact without any approximations.

Second, we transform the Hermitian matrix $({\bf 1} - \eta)$ via the unitary matrix $U_1^{}$ and then obtain
\begin{eqnarray}
	U_1^{} \cdot ({\bf 1} - \eta) = \begin{pmatrix}
		a_1^{} && B_1^{} && B_2^{} \\
		0 && A_{11}^{} && A_{12}^{} \\
		0 && A_{21}^{} && A_{22}^{}
	\end{pmatrix}\;,
	\label{eq:U1(1-eta)}
\end{eqnarray}
where the matrix elements are explicitly given by
\begin{eqnarray}
B_1^{} &\equiv& \frac{1}{a^{}_1} \left(-2\eta_{e\mu}^{} + \eta_{ee}^{} \eta_{e\mu}^{} + \eta_{e\mu}^{} \eta_{\mu\mu}^{} + \eta_{e\tau}^{} \eta_{\mu\tau}^\ast\right) \;, \label{eq:B1} \\
B_2^{} &\equiv& \frac{1}{a^{}_1} \left(-2\eta_{e\tau}^{} + \eta_{ee}^{} \eta_{e\tau}^{} + \eta_{e\tau}^{} \eta_{\tau\tau}^{} + \eta_{e\mu}^{} \eta_{\mu\tau}^{}\right) \;, \label{eq:B2} \\
	A_{11}^{} &\equiv& \frac{1}{a_1^{}} \left[\left|\eta_{e\mu}^{}\right|^2_{} -  \left(1 - \eta_{ee}^{}\right)\left(1 - \eta_{\mu\mu}^{}\right) +
	\frac{\left(1 - \eta_{\mu\mu}^{}\right)\left|\eta_{e\tau}^{}\right|^2_{} + \eta_{e \mu}^\ast \eta_{e \tau}^{} \eta_{\mu\tau}^\ast}{ a^{}_1 - (1 - \eta_{ee}^{} )}\right] \;, \label{eq:A11}\\	
	A_{22}^{} &\equiv& \frac{1}{a_1^{}} \left[\left|\eta_{e\tau}^{}\right|^2_{} - \left(1 - \eta_{ee}^{}\right)\left(1 - \eta_{\tau\tau}^{}\right)+
	\frac{\left(1 - \eta_{\tau\tau}^{}\right)\left|\eta_{e\mu}^{}\right|^2_{} + \eta_{e \mu}^{} \eta_{e \tau}^\ast \eta_{\mu\tau}^{}}{ a_1^{} - (1 - \eta_{ee}^{} ) }\right] \;, \label{eq:A22} \\
	A_{21}^{} &\equiv& \frac{1}{a_1^{}} \left[\eta_{e \mu}^{} \eta_{e \tau}^\ast + \eta_{\mu\tau}^\ast \left(1 - \eta_{ee}^{}\right) -
	\frac{\eta_{e \mu}^{} \eta_{e \tau}^\ast \left(1 - \eta_{\mu\mu}^{}\right) + \left|\eta_{e\mu}^{}\right|^2_{} \eta_{\mu\tau}^\ast }{ a_1^{} - (1 - \eta_{ee}^{} )} \right]\;, \label{eq:A21}\\
	A_{12}^{} &\equiv& \frac{1}{a_1^{}} \left[\eta_{e \mu}^\ast \eta_{e \tau}^{} + \eta_{\mu\tau}^{} \left(1 - \eta_{ee}^{}\right) -
	\frac{\eta_{e \mu}^\ast \eta_{e \tau}^{} \left(1 - \eta_{\tau\tau}^{}\right) + \left|\eta_{e\tau}^{}\right|^2_{} \eta_{\mu\tau}^{} }{ a_1^{} - (1 - \eta_{ee}^{} )} \right]\; .	\label{eq:A12}
\end{eqnarray}
Since the first column vector $a^{}_1{\bf e}^{}_1$ of the new matrix on the right-hand side of Eq.~(\ref{eq:U1(1-eta)}) will not be changed by any unitary transformations in the two-dimensional subspace, we can concentrate on the $3\times 3$ unitary matrix $U^{}_2$ that can be reduced to ${\bf 1}\oplus V^{}_2$, where ${\bf 1}$ stands for the unit matrix in the one-dimensional subspace and $V^{}_2$ for the $2\times 2$ unitary matrix in the orthogonal two-dimensional subspace. Similar to the previous procedure, we denote the two-dimensional vector ${\bf x}^\prime_2 = (A^{}_{11}, A^{}_{21})^{\rm T}$ and the target vector ${\bf y}^\prime_2 = a^{}_2 {\bf e}^\prime_2$ with $a^{}_2 = \left(|A_{11}|^2 + |A_{21}|^2\right)^{1/2} > 0$ and ${\bf e}^\prime_2 \equiv (1, 0)^{\rm T}$, and try to determine $V^{}_2$ such that $V^{}_2 \cdot {\bf x}^\prime_2 = {\bf y}^\prime_2$. This can be achieved as follows. First, define the phase $\psi^{}_1 \equiv \arg \left({\bf x}^{\prime \dagger}_2 \cdot {\bf y}^\prime_2\right)$, i.e. $A_{11}^{} = {\rm e}^{-{\rm i} \psi_1^{}} |A_{11}^{}|$, and construct ${\bm \omega}^{}_2 \equiv {\rm e}^{{\rm i} \psi_1^{}} {\bf x}_2^\prime - {\bf y}_2^\prime$. Then, we get $V_2^{} \equiv {\rm e}^{{\rm i} \psi_1^{}} V_{{\bm \omega}^{}_2}^{}$ with $V_{{\bm \omega}^{}_2}^{} \equiv {\bf 1} - 2 ({\bm \omega}^\dagger_2 \cdot {\bm \omega}_2^{})^{-1} {\bm \omega}_2^{} \cdot {\bm \omega}_2^\dagger$. More explicitly, we have
\begin{eqnarray}
	V_2^{} = \frac{{\rm e}^{{\rm i} \psi_1^{}}}{a_2^{}} \begin{pmatrix} \left|A_{11}^{}\right| && {\rm e}^{-{\rm i} \psi_1^{}} A_{21}^\ast \\ {\rm e}^{{\rm i} \psi_1^{}} A_{21}^{} && - \left|A_{11}^{}\right| \end{pmatrix}\,.
\end{eqnarray}
Now we make a further transformation of the matrix in Eq.~(\ref{eq:U1(1-eta)}) by using the unitary matrix $U_2^{} \equiv {\bf 1} \oplus V_2^{}$, and arrive at
\begin{eqnarray}
	U_2^{} \cdot U_1^{} \cdot ({\bf 1} - \eta) = \begin{pmatrix}
	    a_1^{} && B_1^{} && B_2^{} \\
		0 &&  a_2^{} && \left(A_{11}^\ast A_{12}^{} + A_{21}^\ast A_{22}^{} \right)/a^{}_2 \\
		0 && 0 && A_3^{}
	\end{pmatrix}\;,
\end{eqnarray}
with $A_3^{} \equiv {\rm e}^{2{\rm i} \psi_1^{}} \left(A_{21}^{} A_{12}^{} - A_{11}^{} A_{22}^{}\right)/ a_2^{}$. The last step is to remove the phase $\psi^{}_2 \equiv \arg A^{}_3$ by using the unitary matrix $U_3^{} = {\rm Diag}\{ 1, 1, {\rm e}^{-{\rm i} \psi_2^{}} \}$. Finally, we collect all the three unitary transformations and complete the QR factorization, i.e.,
\begin{eqnarray}
	U_3^{} \cdot U_2^{} \cdot U_1^{} \cdot ({\bf 1} - \eta) = \begin{pmatrix}
		a_1^{} && B_1^{} && B_2^{} \\
		0 && a_2^{} && \left(A_{11}^\ast A_{12}^{} + A_{21}^\ast A_{22}^{} \right)/a^{}_2 \\
		0 && 0 && \left| A_3^{} \right|
	\end{pmatrix} \equiv R \;,
\end{eqnarray}
where $R$ is an upper-triangular matrix with real and positive diagonal elements. Therefore, the QR factorization of the Hermitian matrix $({\bf 1} - \eta)$ is given by ${\bf 1} - \eta = Q \cdot R$, where the unitary matrix reads $Q = U_1^\dagger \cdot U_2^\dagger \cdot U_3^\dagger$. As one can easily verify, the relation ${\bf 1} - \eta = Q \cdot R = R^\dagger_{} \cdot Q^\dagger_{}$ holds.

From the above discussions, it is straightforward to identity $T = R^\dagger_{}$ and $\widetilde{V} = Q^\dagger \cdot V^\prime$ in the triangular parametrization in Eq.~(\ref{eq:trianglepara}), where $V^\prime$ is the unitary matrix involved in the Hermitian parametrization in Eq.~(\ref{eq:hermitianpara}). The nonzero matrix elements of $T$ read
\begin{eqnarray}
	\alpha_{11}^{} &=& a_1^{} = \sqrt{ (1 - \eta_{ee}^{})^2_{} + | \eta_{e \mu}^{} |^2_{} + | \eta_{e \tau}^{} |^2_{} }\;, \\
	\alpha_{22}^{} &=&  a_2^{} = \sqrt{|A_{11}|^2 + |A_{21}|^2}\;,\\
	\alpha_{33}^{} &=& \left|A_3^{}\right| = \left|A_{21}^{} A_{12}^{} - A_{11}^{} A_{22}^{}\right|/ a_2^{}\;,\\
	\alpha_{21}^{} &=& B_1^\ast = \left(-2\eta_{e\mu}^\ast + \eta_{ee}^{} \eta_{e\mu}^\ast + \eta_{e\mu}^\ast \eta_{\mu\mu}^{} + \eta_{e\tau}^\ast \eta_{\mu\tau}^{} \right)/ a_1^{} \;,\\
	\alpha_{31}^{} &=& B_2^\ast = \left(-2\eta_{e\tau}^\ast + \eta_{ee}^{} \eta_{e\tau}^\ast + \eta_{e\tau}^\ast \eta_{\tau\tau}^{} + \eta_{e\mu}^\ast \eta_{\mu\tau}^\ast \right)/ a_1^{} \;,\\
	\alpha_{32}^{} &=& \left(A_{11}^{} A_{12}^\ast + A_{21}^{} A_{22}^\ast \right)/ a_2^{} \; ,
\end{eqnarray}
where the relevant parameters can be found in Eqs.~(\ref{eq:B1})-(\ref{eq:A12}). At the zeroth order of $|\eta^{}_{\alpha \beta}|$, the flavor mixing matrix $V$ is actually unitary. In this case, we shall have $V \approx V^\prime$ in the Hermitian parametrization and $V \approx \widetilde{V}$ in the triangular parametrization, because of ${{\bf 1} - \eta} \approx {\bf 1}$ and $T \approx {\bf 1}$ at this order. In fact, we can directly calculate $Q^\dagger = U^{}_3 \cdot U^{}_2 \cdot U^{}_1$, which approximates to
\begin{eqnarray}
	Q^\dagger_{} \approx \begin{pmatrix}
		1 && 0 && 0 \\ 0 && \cos 2\theta && -\sin 2\theta \\
		0 && -\sin 2\theta && -\cos 2\theta
	\end{pmatrix} \cdot
	\begin{pmatrix}
		1 && 0 && 0 \\  0 && \cos 2\theta && -\sin 2\theta \\
		0 && -\sin 2\theta && -\cos 2\theta
	\end{pmatrix}
	= {\bf 1} \; ,
\end{eqnarray}
where $\sin^2_{}\theta \equiv |\eta_{e \mu}^{}|^2_{} / (|\eta_{e \mu}^{}|^2_{} + |\eta_{e \tau}^{}|^2_{})$ has been defined. To the first order of $|\eta^{}_{\alpha \beta}|$, one obtains
\begin{eqnarray}
	U_1^{} \approx \begin{pmatrix}
		1 && -\eta_{e\mu}^{} && -\eta_{e\tau}^{} \\
		-\eta_{e\mu}^\ast && \cos 2\theta && -\sin 2\theta \; {\rm e}^{-{\rm i}\varphi}_{} \\
		-\eta_{e\tau}^\ast &&  -\sin 2\theta \; {\rm e}^{{\rm i}\varphi }_{} && -\cos 2\theta
	\end{pmatrix}\;,
\end{eqnarray}
with $\varphi \equiv \arg(\eta_{e\mu}^{} \eta_{e\tau}^*)$,
\begin{eqnarray}
	U_2^{} \approx \begin{pmatrix}
		1 && 0 && 0 \\
		0 && 1 && -\eta_{\mu\tau}^{} \\
		0 && -\eta_{\mu\tau}^\ast && -1
	\end{pmatrix} \cdot \begin{pmatrix}
		1 && 0 && 0 \\
		0 && \cos 2\theta && -\sin 2\theta \; {\rm e}^{-{\rm i}\varphi}_{} \\
		0 && -\sin 2\theta \; {\rm e}^{{\rm i}\varphi}_{} && -\cos 2\theta
	\end{pmatrix}\;,
\end{eqnarray}
and $U_3^{} = {\rm Diag}\{1, 1, -1\}$. Thus, at this order, we get
\begin{eqnarray}
	Q^\dagger_{} = U^{}_3 \cdot U^{}_2 \cdot U^{}_1 \approx  \begin{pmatrix}
		1 && -\eta_{e\mu}^{} && -\eta_{e\tau}^{} \\
		+\eta_{e\mu}^\ast && 1 && -\eta_{\mu\tau}^{}\\
		+\eta_{e\tau}^\ast && +\eta_{\mu\tau}^\ast && 1
	\end{pmatrix} \; ,
\end{eqnarray}
which changes the Hermitian matrix $({\bf 1} - \eta)$ into the upper-triangular matrix
\begin{eqnarray}
	T^\dagger_{} = Q^\dagger_{} \cdot ({\bf 1}-\eta) \approx \begin{pmatrix}
		1-\eta_{ee}^{} && -2 \eta_{e\mu}^{} && -2 \eta_{e\tau}^{} \\
		0 && 1-\eta_{\mu\mu}^{} && -2 \eta_{\mu\tau}^{} \\
		0 && 0 && 1-\eta_{\tau\tau}^{}
	\end{pmatrix}\;.
\end{eqnarray}
These are just the results presented in Eqs.~(\ref{eq:RQ}) and (\ref{eq:approxid}).

\section{Jarlskog-like Parameters}
\label{app:UN}
In Refs.~\cite{Xing:2013ty, Xing:2013woa}, the CP asymmetries in neutrino-antineutrino oscillations have been studied and the relevant Jarlskog-like parameters $\widetilde{\mathcal{V}}_{\alpha\beta}^{ij}$ in the case of a unitary mixing matrix have been derived. Since we attempt to make a comparison between the results in the unitary and non-unitary cases, the explicit formulas of the Jarlskog-like parameters, which are calculated by using the unitary mixing matrix $\widetilde{V}$ in Eq.~(\ref{eq:standardpara}), will be collected in this Appendix. As the mixing matrix $\widetilde{V}$ is unitary, one can prove the following identities~\cite{Xing:2013woa}
\begin{eqnarray}
	\widetilde{\mathcal{V}}_{e \mu}^{ij} &=& \frac{1}{2}\left(\widetilde{\mathcal{V}}_{\tau \tau}^{ij} - \widetilde{\mathcal{V}}_{ee}^{ij} - \widetilde{\mathcal{V}}_{\mu \mu}^{ij}\right)\,\label{relaemu},\\
	\widetilde{\mathcal{V}}_{e \tau}^{ij} &=& \frac{1}{2}\left(\widetilde{\mathcal{V}}_{\mu \mu}^{ij} - \widetilde{\mathcal{V}}_{ee}^{ij} - \widetilde{\mathcal{V}}_{\tau \tau}^{ij} \right)\,\label{relaetau},\\
	\widetilde{\mathcal{V}}_{\mu \tau}^{ij} &=& \frac{1}{2}\left(\widetilde{\mathcal{V}}_{ee}^{ij} - \widetilde{\mathcal{V}}_{\mu \mu}^{ij} - \widetilde{\mathcal{V}}_{\tau \tau}^{ij} \right) \, \label{relamutau}\,,
\end{eqnarray}
which together with $\widetilde{\mathcal{V}}_{\alpha \beta}^{ij} = \widetilde{\mathcal{V}}_{\beta \alpha}^{ij} = -\widetilde{\mathcal{V}}_{\alpha \beta}^{ji} = -\widetilde{\mathcal{V}}_{\beta \alpha}^{ji}$ and $\widetilde{\mathcal{V}}_{\alpha\beta}^{ii} = 0$ lead to only nine independent $\widetilde{\mathcal{V}}_{\alpha\alpha}^{ij}$ (for $ij = 12, 13, 23$ and $\alpha = e, \mu, \tau$). In the standard parametrization of $\widetilde{V}$, they can be explicitly written as~\cite{Xing:2013woa}
\begin{eqnarray}
	\widetilde{\mathcal{V}}_{ee}^{12} &=& s_{12}^2 c_{12}^2 c_{13}^4 \sin 2(\rho - \sigma)\,,\\
	\widetilde{\mathcal{V}}_{ee}^{13} &=& c_{12}^2 s_{13}^2 c_{13}^2 \sin 2(\rho + \delta)\,,\\
	\widetilde{\mathcal{V}}_{ee}^{23} &=& s_{12}^2 s_{13}^2 c_{13}^2 \sin 2(\sigma + \delta)\,,\\
\end{eqnarray}
and
\begin{eqnarray}
	\widetilde{\mathcal{V}}_{\mu\mu}^{12} &=& s_{12}^2 c_{12}^2 \left(c_{23}^4 - 4 s_{13}^2 s_{23}^2 c_{23}^2 + s_{13}^4 s_{23}^4\right) \sin 2(\rho - \sigma) + 2 {\cal J}^{}_{12} \left(c_{23}^2 - s_{13}^2 s_{23}^2\right)\nonumber\\
	&& + s_{13}^2 s_{23}^2 c_{23}^2 \left[c_{12}^4 \sin2(\rho - \sigma + \delta) + s_{12}^4 \sin2(\rho - \sigma - \delta)\right],\\
	\widetilde{\mathcal{V}}_{\mu\mu}^{13} &=& c_{13}^2 s_{23}^2 \left[s_{12}^2 c_{23}^2 \sin2\rho + 2{\cal J}^{}_{\rm r} \sin (2\rho + \delta) + c_{12}^2 s_{13}^2 s_{23}^2 \sin 2(\rho + \delta)\right]\,,\\
	\widetilde{\mathcal{V}}_{\mu\mu}^{23} &=& c_{13}^2 s_{23}^2 \left[c_{12}^2 c_{23}^2 \sin2\sigma - 2{\cal J}^{}_{\rm r} \sin (2\sigma + \delta) + s_{12}^2 s_{13}^2 s_{23}^2 \sin 2(\sigma + \delta)\right]\,,
\end{eqnarray}
and
\begin{eqnarray}
	\widetilde{\mathcal{V}}_{\tau\tau}^{12} &=& s_{12}^2 c_{12}^2 \left(s_{23}^4 - 4 s_{13}^2 s_{23}^2 c_{23}^2 + s_{13}^4 c_{23}^4\right) \sin 2(\rho - \sigma) - 2{\cal J}^{}_{12} \left(s_{23}^2 - s_{13}^2 c_{23}^2\right)\nonumber\\
	&& + s_{13}^2 s_{23}^2 c_{23}^2 \left[c_{12}^4 \sin2(\rho - \sigma + \delta) + s_{12}^4 \sin2(\rho - \sigma - \delta)\right],\\
	\widetilde{\mathcal{V}}_{\tau\tau}^{13} &=& c_{13}^2 c_{23}^2 \left[s_{12}^2 s_{23}^2 \sin2\rho - 2{\cal J}^{}_{\rm r} \sin (2\rho + \delta) + c_{12}^2 s_{13}^2 c_{23}^2 \sin 2(\rho + \delta)\right]\,,\\
	\widetilde{\mathcal{V}}_{\tau\tau}^{23} &=& c_{13}^2 c_{23}^2 \left[c_{12}^2 s_{23}^2 \sin2\sigma + 2{\cal J}^{}_{\rm r} \sin (2\sigma + \delta) + s_{12}^2 s_{13}^2 c_{23}^2 \sin 2(\sigma + \delta)\right]\,,
\end{eqnarray}
where the reduced Jarlskog invariant ${\cal J}^{}_{\rm r} \equiv {\cal J}/\sin\delta = s^{}_{12} c^{}_{12} s^{}_{13} c^2_{13} s^{}_{23} c^{}_{23}$ and the definition of ${\cal J}^{}_{12}$ in Eq.~(\ref{eq:J12}) have been used.

\end{document}